\documentclass[reprint,author-numerical,nofootinbib]{revtex4-2}
\usepackage[utf8]{inputenc}
\usepackage{amsmath}
\usepackage{amsfonts}
\usepackage{graphicx}% Include figure files
\usepackage{bm}% bold math
\def\dif{{\rm d}}

\newcommand{\be}{\begin{equation}}
\newcommand{\ee}{\end{equation}}

\begin{document}

\preprint{AIP/123-QED}

\title[Thermodynamic fluid spheres admitting flat synchronization]
{Thermodynamic perfect fluid spheres admitting an orthogonal flat synchronization}

%\\This line break forced with \textbackslash\textbackslash
\author{Salvador Mengual}
\affiliation{ Departament d'Astronomia i Astrof\'{\i}sica,
Universitat de Val\`encia, E-46100 Burjassot, Val\`encia, Spain.}
\author{Joan Josep Ferrando}
\altaffiliation[Also at ]{Observatori Astron\`omic, Universitat
de Val\`encia,  E-46980 Paterna, Val\`encia, Spain}%Lines break automatically or can be forced with \\
\email{joan.ferrando@uv.es.} \affiliation{ Departament d'Astronomia
i Astrof\'{\i}sica, Universitat
de Val\`encia, E-46100 Burjassot, Val\`encia, Spain.}%\\This line break forced with \textbackslash\textbackslash
%\author{Juan Antonio S\'aez}
%\affiliation{ Departament de Matem\`atiques per a l'Economia i l'Empresa,
%Universitat de Val\`encia, E-46022 Val\`encia, Spain.%\\This line break forced% with \\}%

\date{\today}% It is always \today, today,

\begin{abstract}
 We analyze the interpretation of the spherically symmetric perfect fluid solutions that admit a flat synchronization orthogonal to the fluid flow as a thermodynamic perfect fluid in local thermal equilibrium. The ideal gas sonic condition is examined for this family of metrics, and the macroscopic conditions for physical reality are accurately tested for some specific solutions.
\end{abstract}

%Use showkeys class option if keyword
\maketitle

\section{Introduction}
\label{sec-intro}

The Lema\^itre-Tolman (LT) model \cite{[{}][{ [English translation: 1997 {\em Gen. Relativ. Gravit.} {\bf 29} 641]}]Lemaitre, [{}][{ [English translation: 1997 {\em Gen. Relativ. Gravit.} {\bf 29} 935]}]Tolman} is a remarkable spherically symmetric dust solution for modeling both gravitational collapse and cosmological inhomogeneities (see also references \cite{[{}][{ [English translation: 1999 {\em Gen. Relativ. Gravit.} {\bf 31} 1783]}]Bondi, kramer, Krasinski, Krasinski-Plebanski, Ellis-Maartens-MacCallum}).  

In the last three decades a large number of studies have been devoted to analyzing the cosmic censorship conjecture by using the LT models (see \cite{Krasinski, Krasinski-Plebanski, Lapi-Morales, Mosani-J} and references therein). In cosmology, the LT solutions provide exact inhomogeneous models for studying the formation of structures \cite{Krasinski, Krasinski-Plebanski, Ellis-Maartens-MacCallum, Krasinski-Hellaby} and for analyzing the effect of the non-linear inhomogeneities on the cosmic microwave background radiation \cite{Krasinski, Krasinski-Plebanski, SDiego1, SDiego2}.

Interest in LT cosmological models grew when several studies revealed that large-scale spatial inhomogeneities of the Universe could be compatible with the cosmological observations (see, for example \cite{Mustapha}), and that the setting of the magnitude-redshift relation with the Type Ia supernovae data can be carried out in an inhomogeneous model without cosmological constant \cite{Celerier, Iguchi, Celerier-Krasinski}. Despite the fact that some authors are skeptical that inhomogeneous models can be compatible with all the cosmological observations (see \cite{Ellis-Maartens-MacCallum} and references therein), several works carry on investigating this subject \cite{Krasinski-Hellaby-B-C, Krasinski_14a, Krasinski_14b}, and it is still an open question.

Although Lema\^itre \cite{Lemaitre} also considered a non-null pressure in his pioneering paper, the dust model is the only one contemplated in the above cited references and in most approaches where the LT metric is considered. Nevertheless, in  some papers the role of pressure is analyzed \cite{Lasky-Lun, Lynden-Bicak} (see also references therein), and some models with anisotropic pressure have been considered \cite{Sussman-98, Sussman-Pavon}. However, something is still lacking in the study of these LT metrics with pressure: their interpretation as a thermodynamic perfect fluid in local thermal equilibrium. 

Note that the study of the physical interpretation of the formal perfect fluid solutions to the Einstein field equations is an open problem in theoretical relativity. Many of these solutions have been obtained by imposing constraints that simplify the integration of the field equations, and very few solutions have been interpreted as physically realistic fluids. A first step in this task consists of analyzing the necessary macroscopic constraints for physical reality and of performing a suitable method to examine whether a specific family of perfect fluid solutions fulfills them.

If we look for solutions that represent perfect fluids in local thermal equilibrium, in addition to imposing the energy conditions \cite{Plebanski}, we must  add to the {\em hydrodynamic} quantities that appear in the energy tensor $T = (\rho+ p) u \otimes u + p \, g$, a set of {\em thermodynamic} quantities constrained by the usual thermodynamic laws \cite{Eckart}. Moreover, in order to obtain a coherent theory of shock waves, the relativistic compressibility conditions \cite{Israel, Lichnero-1} must be required. Our macroscopic hydrodynamic approach to the local thermal equilibrium \cite{CFS-LTE} and to the relativistic compressibility conditions \cite{CFS-CC} provides a tool to impose these physical requirements. 

Based on this essential groundwork we have studied the ideal gas Stephani universes \cite{CFS-CC, CF-Stephani}, the classical ideal gas solutions \cite{CFS-CIG}, and the thermodynamic Szekeres-Szafron solutions \cite{FS-SS, CFS-PSS, CFS-RSS}. Currently, we plan to study the physical meaning of the perfect fluid solutions admitting a group G$_3$ of isometries acting on spacelike two-dimensional orbits S$_2$. Our recent results on the thermodynamic T-models \cite{FM-Tmodels, FM-Tmodels2} are on the way to achieving this goal. The present paper is the first step to undertake a similar study for the R-models (the curvature of the orbits S$_2$ has a gradient that is not tangent to the fluid flow).  

The R-models with geodesic motion are the perfect fluid solutions whose metric line element is given by \cite{Krasinski-Plebanski}:
\begin{equation} \label{metric-Y-E}
ds^2= -dt^2 + \frac{[Y'(t,r)]^2}{1+ 2 E(r)} \, dr^2 + Y^2(t,r) \, d\tilde{\Omega}^2,
\end{equation}
where $d\tilde{\Omega}^2$ is a two-dimensional metric of constant curvature, and a prime represents partial derivative with respect to the coordinate $r$. 

The commonly considered LT models are the metrics of the form (\ref{metric-Y-E}) ({\em LT metrics}) with zero pressure and cosmological constant. Although very few explicit solutions with non-constant pressure are known, the perfect fluid general solution can be obtained by quadratures for the spatially flat case $E(r)=0$ \cite{Bona-Stela-a}. This result made it possible to construct "Swiss cheese" cosmological models with pressure \cite{Bona-Stela-c}.

The procedure that we have developed in the above quoted papers to analyze the physical meaning of the perfect fluid solutions (see \cite{CFS-RSS, FM-Tmodels} for more details) could be applied to the full set of the perfect fluid solutions of the form (\ref{metric-Y-E}). In this case, the obtained constraints would be simply formal and of little practical interest. Nevertheless, when we apply our procedure to a family of explicit solutions, we can go further in our analysis of the physical meaning of the solutions. Therefore, we will limit ourselves here to studying the case in which explicit solutions are known, that is, when $E(r)=0$.

Two comments about this spatially flat case. Firstly, only the spherical symmetry is compatible \cite{kramer}, and consequently $d\tilde{\Omega}^2= d \Omega^2$ is the metric of a two-sphere. Secondly, the spherically symmetric perfect fluid solutions that admit a flat slice orthogonal to the fluid flow have a geodesic motion \cite{Bona-Stela-a}, and thus they coincide with the perfect fluid solutions for the LT metrics (\ref{metric-Y-E}) with $E(r)=0$.

Thus, here we study the thermodynamics and analyze the macroscopic conditions for physical reality of the spherically symmetric perfect fluid solutions admitting a flat synchronization orthogonal to the fluid flow. 

In Sec. \ref{sec-metric} we present the metric line element and remark on the linearity of the field equations. Several approaches that can be considered in solving them are also sketched. As an example we obtain the general solution of the flat dust LT models and we comment on the previously known results.

In Sec. \ref{sec-termo} we undertake the general study of the thermodynamics of the solutions. On the one hand, we obtain the kinematic and hydrodynamic quantities (expansion, pressure and energy density of the fluid) as well as the indicatrix function that gives the square of the speed of sound. On the other hand, we determine the thermodynamic schemes that are compatible with each model. 

If we want to go further in our analysis of the physical meaning of the solutions we must specify the general expressions presented in the previous section by considering particular solutions or by adding some physical properties. In Sec. \ref{sec-chi-pi} we impose a significant physical constraint on the models; their compatibility with the equation of state of a generic ideal gas. The sonic condition and the other macroscopic physical requirements are specified for this ideal case, and the general equations that characterize these {\em ideal models} are obtained.

In Sec. \ref{sec-t^q} we analyze when the spherically symmetric limit of the Szafron solution \cite{Szafron} fulfills the ideal constraints studied above. The behavior of the subsequent {\em ideal Szafron models} is accurately analyzed, and the spacetime domains where the macroscopic constraints for physical reality hold are obtained. A similar study is undertaken in Sec. \ref{sec-t^1/2} for another ideal model, which can be considered as a limit of the ideal Szafron models. 

In Sec. \ref{sec-progress} we give a preview of several results on some open topics. First, we study the conditions that characterize the models consistent with a homogeneous temperature, namely, those compatible with a non-vanishing thermal conductivity coefficient. Second, we outline with an example how to determine new solutions to the ideal sonic equation obtained in Sec. \ref{sec-chi-pi}.

%%%%%%%%%%

\section{Metric and general solution}
\label{sec-metric}

In synchronous comoving coordinates the metric of the spherically symmetric perfect fluid solutions that admit a flat slice orthogonal to de fluid flow is \cite{Krasinski-Plebanski}
\begin{equation} \label{metric-Y}
ds^2= -dt^2 + [Y'(t,r)]^2 \, dr^2 + Y^2(t,r) \, d\Omega^2,
\end{equation}

The unit velocity of the fluid is $u = \partial_t$, and its expansion and the non-vanishing components of the shear tensor are, respectively:
\begin{eqnarray} \label{expansion-Y}
\theta = 2 \, \frac{\dot{Y}}{Y}+\frac{\dot{Y}'}{Y'} = \partial_t [\ln(Y^2 \, Y')] ,\\[1mm]
%\end{equation}
%
%
%\begin{equation}
 \label{shear-Y}
\sigma^1_1 = \sigma^2_2 = -\frac{1}{2}\sigma^3_3 = \bar{\sigma} \equiv  \frac{1}{3}\left( \frac{\dot{Y}}{Y}-\frac{\dot{Y}'}{Y'} \right)  ,
\end{eqnarray}
where a dot represents partial derivative with respect to the time coordinate $t$. Note that, $u$ being geodesic, we have $u(\phi) \equiv u^{\alpha} \partial_{\alpha} \phi = \dot{\phi}$, for any spacetime function $\phi(x^{\alpha})$.

The Einstein equations for a perfect fluid source reduce to the following expressions for the \textit{pressure} $p$ and the \textit{energy density} $\rho$ \cite{kramer}
\begin{equation} \label{pressure and density-Y}
p = -2 \, \frac{\ddot{Y}}{Y} - \frac{\dot{Y}^2}{Y^2} \, ,\quad\,  \quad\, \rho =  2 \, \frac{\dot{Y}}{Y}\frac{\dot{Y}'}{Y'} + \frac{\dot{Y}^2}{Y^2} \, ,
\end{equation}
where $p=p(t)$ as a consequence of the conservation of the energy tensor, $\nabla \cdot T = 0$. If we do not add any additional physical requirement only the first equation in (\ref{pressure and density-Y}) constrains the metric function $Y(t,r)$, and the second one gives the energy density for a given solution.

\subsection{General solution of the field equations} 
\label{subsec-metric-solucio}

By performing the substitution
\begin{equation} \label{def. Z}
Y=Z^{2/3} \, ,
\end{equation}
expressions (\ref{expansion-Y}), (\ref{shear-Y}), and (\ref{pressure and density-Y}) for the expansion, shear, pressure, and energy density become
\begin{eqnarray} \label{expansion-shear-Z}
\theta = \frac{\dot{Z}}{Z} + \frac{\dot{Z}'}{Z'} \, , \qquad \bar{\sigma} = \frac{1}{3}\left( \frac{\dot{Z}}{Z}-\frac{\dot{Z}'}{Z'}\right) , \\
%\end{equation}
%
%\begin{equation} 
\label{pressure and density-Z}
p = -\frac{4}{3} \, \frac{\ddot{Z}}{Z} \, ,\qquad \quad  \rho  = \frac{4}{3}\, \frac{\dot{Z}}{Z}\, \frac{\dot{Z}'}{Z'} \, .
\end{eqnarray}
Then, from the expression for the pressure we obtain the following equation \cite{Bona-Stela-a}
\begin{equation} \label{eq. Z}
\ddot{Z} + \frac{3}{4}p(t)Z = 0 \, ,
\end{equation}
which is linear in Z. Thus, the general solution is of the form \cite{Bona-Stela-a}
\begin{equation} \label{sol. general eq. Z}
Z = a(r) \, f(t) + b(r) \, g(t) \, ,
\end{equation}
where $a(r)$ and $b(r)$ are arbitrary functions of the radial coordinate $r$, and $f(t)$ and $g(t)$ are two independent particular solutions of (\ref{eq. Z}).

Therefore, the problem of finding particular solutions of the field equations can be tackled following different approaches:

(i) On the one hand, we can give an arbitrary pressure $p(t)$ and look for the general solution to (\ref{eq. Z}). Note that the particular solutions $f$ and $g$ to equation (\ref{eq. Z}) are related by 
\begin{equation} \label{gtt(f)}
\ddot{f}/f = \ddot{g}/g \, ,
\end{equation}
which can be integrated to give $\dot{g} f= \dot{f} g + C$, where $C$ is an arbitrary integration constant. However, since we only need $g$ to be any non-trivial particular solution of (\ref{eq. Z}) independent to $f$, we can set $C=1$ and therefore obtain 
\begin{equation} \label{gt(f)}
\dot{g} f= \dot{f} g + 1 \, .
\end{equation}
This allows us to obtain from a known particular solution $f(t)$, another particular solution $g(t)$ as  \cite{Bona-Stela-a, kramer}
\begin{equation} \label{g(f)}
g(t) = f(t)\! \int^t \! \! \! f^{-2}(t') \, dt' \, .
\end{equation}
For instance, for $p = 0 $ Eq. (\ref{eq. Z}) can easily be solved to obtain $Z(t,r) = a(r) \, t + b(r)$, which corresponds to the parabolic subset ($E(r)=0$) of the Lema\^itre-Tolmann dust models \cite{Bona-Stela-a, Krasinski-Plebanski}.

(ii) On the other hand, we can give an arbitrary function $f(t)$ as input, and then we can determine the pressure as $3 p = -4 \ddot{f}/f$, and complete the solution of (\ref{eq. Z}) by using (\ref{g(f)}). Thus, we can obtain the solution to the field equations by quadratures. For instance,  by choosing $f(t) = t^q$, $q \neq 1/2$, then $p \sim t^{-2}$ and $g(t)= t^{1-q}/(1-2q)$. These solutions correspond to the spherically symmetric subset of a wider family of solutions considered by Szafron \cite{Szafron}.

(iii) However, one could also start by making a particular election of an arbitrary function $\varphi(t)$ such that $\dot{\varphi}(t)>0$, and then obtain $f(t)$ and $g(t)$ as \cite{Bona-Stela-b}
\be
f(t) = [\dot{\varphi}(t)]^{-1/2} \, , \qquad g(t) = f(t)  \varphi(t) \, .
\ee
Note that this approach allows us to obtain the general solutions of the field equations without the need to calculate any integrals. 

It is worth remarking that the same procedures can be carried out for the flat class I Szekeres-Szafron solutions since Eq. (\ref{eq. Z}) is also fulfilled in that case \cite{Bona-Stela-b, Szafron}. However, although our study of the thermodynamic properties is based on the solution of Eq. (\ref{eq. Z}), our results only apply in spherical symmetry since the class I Szekeres-Szafron solutions only admit a thermodynamic interpretation in this case \cite{KQS}.

Notice that a redefinition of the radial coordinate $r$ allows us to consider one of the functions $a(r)$ or $b(r)$ to be any real function. In fact the metric only depends on the quotient $\alpha = \alpha(r) \equiv a(r)/b(r)$. Consequently the degrees of freedom of the spherically symmetric perfect fluid solutions admitting a flat synchronization are given by the election of an arbitrary function of $r$, say $\alpha(r)$, and an arbitrary function of time, either $f(t)$ or $p(t)$ or $\varphi(t)$. So, the gravitational field is determined by a pair $\{f(t), \alpha(r)\}$.

The spatially homogeneous limit of these solutions are the flat Friedmann-Lema\^itre-Robertson-Walker (FLRW) metrics, which can be characterized by one of the following five equivalent conditions: (i) the metric function $Y(t,r)$ factorizes, and then the coordinate $r$ can be taken so that $Y(t,r) = r R(t)$; (ii) $\alpha(r) = a(r)/b(r)$ is a constant function; (iii)  the fluid expansion is homogeneous, $\theta = \theta(t)$; (iv) the fluid flow is shear-free, $\bar{\sigma} = 0$; and (v) the energy density is homogeneous $\rho= \rho(t)$, and then the fluid evolution is barotropic, $d \rho \wedge d p =0$.

%%%%%%%%%%%%%%%%%%%

\subsection{On the dust flat LT models}
\label{subsec-pressure constant}

As an example, in this subsection we will consider the dust flat LT models, that is, the solutions in which the pressure $p$ takes a constant value, $p = -\Lambda$ \cite{Krasinski-Plebanski, barrow}. Now Eq. (\ref{eq. Z}) becomes 
\begin{equation} \label{eq:Z-Lambda}
\ddot{Z} - \frac34 \Lambda \; Z = 0 \, ,
\end{equation}
and the energy density is $\rho = \Lambda + \rho_H$, where $\rho_H$ is the hydrodynamic energy density ($\equiv$ matter density).  
The case $\Lambda=0$ is the most frequently considered in the literature \cite{Krasinski-Plebanski, Krasinski}, and the solution can be written as
\be \label{Z-Lambda=0}
Z = Z_0(r) [t-t_0(r)] \, , 
\ee
where $t_0(r)$ is the non-simultaneous big bang time, and $M(r) = (2/9) Z_0^2(r)$ is the effective gravitational mass. Moreover, the matter density takes the expression
\be
\rho_H  = \frac{4}{3(t\! -\! t_0)^2 \left[1- \frac{Z_0\,  t_0'}{Z_0' (t-t_0)}\right]} \, .
\ee

These expressions can be generalized for $\Lambda \not=0$. Indeed, the general solution to (\ref{eq:Z-Lambda}) is
\begin{equation} 
\hspace*{-2.0mm}
Z(t,r)\! = \!\begin{cases}
 a(r) \sinh \, \omega t + b(r) \cosh \, \omega t   , \quad \textrm{if} \;\; \Lambda > 0 \cr 
 a(r) \sin \, \omega t + b(r) \cos \, \omega t   , \quad\quad\, \textrm{if} \;\; \Lambda < 0 
 \end{cases}
\end{equation}
where $\omega^2 \equiv \frac34 |\Lambda|$. It is worth remarking that, for the $\Lambda > 0$ case, if $a(r) > b(r)$ (respectively, $b(r) > a(r)$) these arbitrary functions can be written as $a(r) = Z_0(r) \cosh[\omega t_0(r)]$ and $b(r) = -Z_0(r) \sinh[\omega t_0(r)]$ (respectively, $b(r) = Z_0(r) \cosh[\omega t_0(r)]$ and $a(r) = -Z_0(r) \sinh[\omega t_0(r)]$). Consequently, we obtain that the general solution $Z(t,r)$ of Eq. (\ref{eq:Z-Lambda}) for $\Lambda > 0$ leads to two models,
\begin{eqnarray} \label{Z-Lambda>0-a}
Z(t,r) = 
 Z_0(r) \sinh (\omega [t\! -\! t_0(r)])   ,  \\[1mm]
Z(t,r) =  Z_0(r) \cosh (\omega [t\! -\! t_0(r)])   .  \label{Z-Lambda>0-b}
\end{eqnarray}
For the model (\ref{Z-Lambda>0-a}) $M(r) = (\Lambda/6) Z_0^2(r)$, and the matter density is
\be \label{hydro-rho}
%\hspace{-0.0mm}
\rho_H = \frac{\Lambda}{\sinh [\omega(t\!- \!t_0)]\left[1\!-\! \frac{\omega Z_0 \, t_0'}{Z_0' \coth [\omega(t-t_0)]}\right]}  .
\ee
We obtain a similar expression for case (\ref{Z-Lambda>0-b}), which follows by changing $\Lambda \rightarrow - \Lambda$ and $\sinh \rightarrow \cosh$.

On the other hand, for case $\Lambda < 0$ the arbitrary functions can always be written as $a(r) = Z_0(r) \cos[\omega t_0(r)]$ and $b(r) = - Z_0(r) \sin[\omega t_0(r)]$. Consequently, the general solution $Z(t,r)$ of Eq. (\ref{eq:Z-Lambda}) for $\Lambda < 0$ becomes
\begin{equation} \label{Z-Lambda<0}
Z(t,r) = Z_0(r) \sin(\omega[t - t_0(r)]).
\end{equation}
Now $M(r) = (|\Lambda|/6) Z_0^2(r)$, and the expression of the hydrodynamic energy density is like (\ref{hydro-rho}) with the changes $\Lambda \rightarrow  |\Lambda|$ and $(\sinh, \coth) \rightarrow (\sin, \cot)$. 

Note that the FLRW homogeneous limit follows by considering $t_0(r) = constant$. In this homogeneous case the matter density is positive everywhere for $\Lambda =0$, for model (\ref{Z-Lambda>0-a}), and for model (\ref{Z-Lambda<0}), and negative for model (\ref{Z-Lambda>0-b}). Nevertheless, shell-crossing singularities \cite{Krasinski-Plebanski} could exist in the inhomogeneous models, which disconnect spacetime domains with positive and negative matter density.

It is worth remarking that the flat dust LT models with $\Lambda=0$ given in (\ref{Z-Lambda=0}) are the most commonly considered in the basic cosmology books \cite{Krasinski, Krasinski-Plebanski, Ellis-Maartens-MacCallum}. On the other hand, the homogeneous limit of solution (\ref{Z-Lambda>0-a}) is the background universe in the standard $\Lambda$CDM cosmological models. This analytical expression is little used by observational and numerical cosmologists, although it was already considered for the first time by Lema\^itre \cite{Lemaitre}, and its generalization to a $\gamma$-law is also known \cite{kramer, Harrison}. 
%
%Wainwright, J. and Ellis, G.F.R. (1997). Dynamical systems in cosmology (Cambridge University Press, Cambridge)
%
The inhomogeneous model (\ref{Z-Lambda>0-a}) has also been considered previously \cite{barrow}.

%%%%%%%%%%%%%%%%%%%%%%

%%%%%%%%%%%%%%%%%%%%%%

\section{Thermodynamics of the solutions}
\label{sec-termo}

If we want a perfect energy-momentum tensor $T$ to describe the energetic evolution of a thermodynamic perfect fluid in \textit{local thermal equilibrium}, we must add to the {\em hydrodynamic quantities} $\lbrace u, \rho, p \rbrace$ a set of {\em thermodynamic quantities} $\lbrace n, s, \Theta \rbrace$ (\textit{matter density} $n$, \textit{specific entropy} $s$ and \textit{temperature} $\Theta$) constrained by the usual thermodynamic laws \cite{Eckart}. Namely, the conservation of matter, 
\begin{equation} \label{conservacio massa}
\nabla \cdot (nu) = u(n) + n \theta = 0 \, ,
\end{equation}
and the {\em local thermal equilibrium relation}, which can be written as
\begin{equation} \label{re-termo}
\Theta \dif s = \dif h - \frac{1}{n} \dif p \, ,  \qquad h \equiv \frac{\rho+p}{n} \, ,
\end{equation}
where $h$ is the {\em relativistic specific enthalpy}.
%; and the decomposition defining the specific internal energy
%
%\begin{equation} \label{masa-energia} 
%\rho= n(1+\epsilon) \, .
%\end{equation}

In \cite{CFS-LTE} we have offered a hydrodynamic condition that guarantees the existence of thermodynamics; a perfect energy-momentum tensor describes the (non isoenergetic, $\dot{\rho} \neq 0$) evolution of a thermodynamic perfect fluid in local thermal equilibrium if, and only if, the hydrodynamic quantities, $\lbrace u, \rho, p \rbrace$, fulfill the so-called \textit{hydrodynamic sonic condition} \cite{CFS-LTE, CFS-CC} 
\begin{equation} \label{sonic-condition}
\hspace{0mm} {\rm S} :  \qquad \quad     \dif \chi \wedge \dif p \wedge \dif \rho = 0 \, , \qquad \chi \equiv \frac{u(p)}{u(\rho)}   \, ,
\end{equation}
where $\chi$ is the \textit{indicatrix of the local thermal equilibrium}. When this condition holds $\chi$ is a function of state, $\chi = \chi(\rho, p)$, which represents the square of the speed of sound in the fluid, $\chi(\rho, p) \equiv c_s^2$. 

Note that due to the symmetries of the metric (\ref{metric-Y}), all scalar invariants depend on two functions at most. Then, the hydrodynamic sonic condition S is fulfilled automatically.

%%%%%%%%%%%%%%%%

\subsection{Hydrodynamic quantities: energy density, pressure and indicatrix function} 
\label{subsec-Termo-hydrodynamic}

Let us consider an inhomogeneous solution of the field equations $Z = a(r) f(t) + b(r) g(t)$, $\alpha'(r) \not=0$, $\alpha \equiv a(r)/b(r)$. If $\beta = \beta(r) \equiv a'(r)/b'(r)$, the fluid expansion (\ref{expansion-shear-Z}) can be written as
\begin{equation} \label{expansion-Zb}
\theta = \partial_t (\ln [(\alpha  f \!+ \! g)(\beta f \!+ \!g)]) = 
\frac{\alpha  \dot{f} + \dot{g}}{\alpha  f+ g} + \frac{\beta  \dot{f} + \dot{g}}{\beta  f+ g} ,
\end{equation}
and the pressure and energy density (\ref{pressure and density-Z}) take the following expressions
\begin{equation} \label{pressure and density-Zb}
p = -\frac{4}{3}\frac{\ddot{f}}{f} \, , \quad \rho = \frac{4}{3} \, \frac{\alpha \beta \dot{f}^2 + (\alpha \!+\! \beta) \dot{f}\dot{g} +\dot{g}^2} {\alpha \beta f^2 + (\alpha \!+\! \beta) fg +g^2} \, .
\end{equation}

The square of the speed of sound can be obtained using the definition (\ref{sonic-condition}) of the indicatrix function, expression (\ref{pressure and density-Zb}) of the pressure, and the energy conservation condition $\dot{\rho} + (\rho+p) \theta =0$,
\begin{equation} \label{chi-general}
c_s^2 = \frac{\dot{p}}{\dot{\rho}} =-\frac{\dot{p}}{\theta (\rho + p)} \equiv  \chi (t,r) \, ,
\end{equation}
where $\theta(t,r)$, $p(t)$, and $\rho(t,r)$ are given in (\ref{expansion-Zb}) and (\ref{pressure and density-Zb}).

If we consider a specific solution of the field equations we can know the expansion $\theta(t,r)$, the pressure $p(t)$, the energy density $\rho(t,r)$ and the indicatrix function $\chi(t,r)$ as explicit spacetime functions, and then we could analyze the physical behavior of the solutions. In particular, we could study the spacetime regions where the energy conditions or the relativistic compressibility conditions hold. It is worth remarking that each solution represents a specific evolution of a family of fluids. If we are interested in the thermodynamic properties of these fluids regardless of the evolution, we should obtain the explicit dependence on the energetic variables of the indicatrix function and thus determine the equation of state $c_s^2 = \chi(\rho,p)$.

%%%%%%%%%%%%%%%%%%%%%%

\subsection{Thermodynamic scheme: Entropy, matter density, and temperature} 
\label{subsec-Termo-scheme}

When a perfect energy tensor $T \equiv \lbrace u, \rho, p \rbrace$ fulfills the sonic condition S, a set of associated thermodynamic quantities $\lbrace n, s, \Theta\rbrace$ exists. This {\em thermodynamic scheme} is not unique. In \cite{CFS-LTE} we have shown that the specific entropies $s$ and the matter densities $n$ associated with $T$ are of the form $s = s(\bar{s})$ and $n = \bar{n}/N(\bar{s})$, where $s(\bar{s})$ and $N(\bar{s})$ are arbitrary real functions of a particular solution $\bar{s} = \bar{s}(\rho, p)$ to the equation $u(s)=0$, and $\bar{n} = \bar{n}(\rho,p)$ is a particular solution to Eq. (\ref{conservacio massa}). Moreover, $\Theta$ is determined by (\ref{re-termo}). Finding the expressions of these thermodynamic quantities corresponds to solving the inverse problem \cite{CFS-LTE} for our metrics (\ref{metric-Y}).

It is worth remarking that only the sonic condition S constrains the metric tensor as a consequence of the Einstein equations, and each associated thermodynamic scheme provides a physical interpretation of the solution; a thermodynamic solution represents the evolution of the family of fluids defined by the thermodynamic properties of the associated schemes.

From expression (\ref{expansion-Zb}) of the expansion it is easy to see that $\bar{n} = [(\alpha  f \!+ \! g)(\beta f \!+ \!g)]^{-1}$ is a particular solution of Eq. (\ref{conservacio massa}), and any function of the radial coordinate $\bar{s}=\bar{s}(r)$ fulfills $u(s)= \dot{s} =0$. Then, we have that the thermodynamic schemes associated with the perfect fluid solutions of the form (\ref{metric-Y}) are determined by a specific entropy $s$ given by $s(\rho, p) = s(r)$, and a matter density $n$ of the form
\begin{equation} \label{n-general}
n(\rho,p) = \frac{1}{N(r)[\alpha \beta f^2 + (\alpha \!+\! \beta)   fg +g^2]}  \, ,
\end{equation}
where $s= s(r)$ and $N = N(r)$ are two arbitrary real functions.

The temperature $\Theta$ of each thermodynamic scheme determined by a pair $\lbrace s, \, n \rbrace$ can be obtained from (\ref{re-termo}) as $\Theta = \left(\frac{\partial h}{\partial s}\right)_p = \frac{1}{s'(r)} \left(\frac{\partial h}{\partial r}\right)_t$. Using (\ref{pressure and density-Zb}) and (\ref{n-general}) we obtain that the specific enthalpy takes the following expression,
\begin{equation} \label{h-general}
%\hspace{-0mm}
h  \! = \!  \frac43 N \!  \left[ \dot{g}^2 \!  - \!  g^2 \ddot{f}/f \!  - \!  \alpha\beta f^2  (\dot{f}/f\dot{)}   \!  \! - \!  (\alpha \! + \! \beta) g^2  (\dot{f}/g\dot{)}  \right] \! .
\end{equation}
Then, we obtain that the temperature $\Theta$ associated with the thermodynamic scheme defined by the functions $\lbrace s(r), N(r) \rbrace$ has the following expression,
\begin{subequations} \label{temperatura-general}
\begin{eqnarray}
%	\displaystyle 
	\Theta = t_1 (t)\, \tau_1 (r) +  t_2 (t)\, \tau_2 (r) +  t_3 (t)\, \tau_3 (r) , \quad  \qquad  \label{temperatura-general-a}\\[1mm]
	 t_1(t) \equiv  \dot{g}^2 \!  - \!  g^2 \ddot{f}/f , \qquad  \tau_1(r) \equiv \frac34 N'/s' ,  \qquad  \qquad  \qquad  \\[-0mm]
	 t_2(t) \equiv  -  f^2    [\dot{f}/f\dot{]\,}   , \quad \quad \, \tau_2(r) \equiv \frac34 (N \alpha \beta)'/s' , \qquad   \qquad    \\[0mm]  
	 t_3(t) \equiv  -  g^2    [\dot{f}/g\dot{]\,} , \quad  \quad \ \tau_3(r) \equiv \frac34 [N (\alpha + \beta)]'/s' .  \quad \quad \ 
\end{eqnarray}
\end{subequations}
%

%%%%%%%%%%%%%%%%%%%%%%

\subsection{Imposing additional physical properties} 
\label{subsec-Termo-MPC}

If we want to go further in our analysis of the physical meaning of the solutions we must impose complementary physical properties on the solutions, and thus specify the general expressions presented in this section. Afterwards, we will be able to impose the macroscopic constraints for physical reality. Among others, we can consider the following approaches:

(i) We can specify a solution by providing the function $f(t)$ that determines the time evolution of the model. For example, we can consider the Szafron solution \cite{Szafron} by taking $f(t)= t^q$ (see Sec. \ref{sec-t^q}).

(ii) We can impose physical constraints on the indicatrix function that fixes the (hydrodynamic) equation of state $c_s^2=\chi(\rho, p)$. For example, we can establish that $\chi(\rho,p)$ is that of a generic ideal gas (see next section). 

(iii) We can impose physical constraints on the thermodynamic scheme. For example, we can demand a homogeneous temperature (see Sec. \ref{subsec-T(t)})
  
 %%%%%%%%%%%%%%%%%%%%%

\section{Ideal Models}
\label{sec-chi-pi}

Now we will analyze when the models considered above are compatible with the equation of state of a {\em generic ideal gas}, namely,
\begin{equation}
p = \tilde{k} n \Theta  \, , \qquad \quad    \tilde{k} \equiv {k_B/m} \,  .  \label{gas-ideal}
\end{equation}
In \cite{CFS-LTE} we have shown that Eq. (\ref{gas-ideal}) restricts the functional dependence of the indicatrix function $c_s^2   = \chi(\rho,p)$. More precisely, a perfect energy tensor $T\equiv \{u,\rho,p\}$ represents the evolution of a generic ideal gas in local thermal equilibrium if, and only if, it fulfills the {\em ideal gas sonic condition},
\be \label{chi-gas-ideal}
\hspace{2mm} {\rm S^{\rm G}} :  \quad \quad     \chi = \chi(\pi) \not= \pi  , \quad \chi \equiv \frac{u(p)}{u(\rho)}  , \quad \pi \equiv \frac{p}{\rho}  .
\ee

On the other hand, in \cite{CFS-CC} we have proved that, for an indicatrix function of the form (\ref{chi-gas-ideal}), $\chi = \chi(\pi)$, the compressibility conditions that constrain the hydrodynamic evolution of the fluid can be written as
\begin{equation}  \label{cc-ideal}
\hspace{2mm} {\rm H}_1^{\rm G} : \quad 
\begin{array}{c}      
0 < \chi < 1  ,  \qquad \quad  \\[2mm] 
\zeta \equiv (1+\pi)(\chi-\pi) \chi'  + 2 \chi(1-\chi) > 0  .  \
\end{array}
\end{equation}
And the remaining compressibility conditions constrain the associated thermodynamic schemes $\{s,n,\Theta\}$ \cite{CFS-CC}
\be \label{H2-Theta}
\hspace{-30mm} {\rm H}_2 : \qquad \qquad  2 n \Theta > \frac{1}{s_{\rho}'}    \, .
\ee
These thermodynamic variables are also bounded by the positivity conditions,
\begin{equation} \label{P}
\hspace{-10mm} {\rm P} :  \qquad \qquad     \Theta > 0 \, , \qquad  \quad   \rho > n > 0   \, .
\end{equation}
Moreover, the equation of state (\ref{gas-ideal}) and the positivity conditions P imply a positive thermodynamic pressure, $p > 0$. Consequently, the energy conditions \cite{Plebanski} for a perfect fluid energy tensor, $ -\rho < p \leq \rho$, become (here we shall consider non-shift perfect fluids, $\rho \not=p$),
\begin{equation} \label{e-c-G}
\hspace{2mm} {\rm E}^{\rm G} : \qquad     \rho > 0 \, , \qquad  0 < \pi < 1 \, , \quad  \pi = p/\rho \, .
\end{equation}

Note that, in order to study the solutions with the hydrodynamic behavior of a generic ideal gas, in a first step we must impose the ideal sonic condition S$^{\rm G}$. In a second step, we must impose on the subfamily thus obtained the hydrodynamic conditions E$^{\rm G}$ and H$_1^{\rm G}$, and determine the spacetime domains where the solution fulfills them. Finally, we must analyze the thermodynamic schemes that are compatible with the constraints P and H$_2$.     

%%%%%%%%%%%%%%

\subsection{Study of the ideal sonic condition {\rm S}$^{\rm G}$}
\label{subsec-ideal-equations}

To obtain the models that meet the ideal sonic condition S$^{\rm G}$ given in (\ref{chi-gas-ideal}) we must impose the constraint $d \chi \wedge d \pi=0$, where $\chi = \chi(t,r)$ is given in (\ref{chi-general}) and where, from (\ref{pressure and density-Zb}), $\pi$ takes the expression
\be
\pi = \frac{p}{\rho} = \pi(t,r) \equiv - \frac{\ddot{f}[\alpha \beta f^2 + (\alpha \!+\! \beta) fg +g^2]}{f[\alpha \beta \dot{f}^2 + (\alpha \!+\! \beta) \dot{f}\dot{g} +\dot{g}^2]} \, .
\ee
Then, if from (\ref{gtt(f)}) and (\ref{gt(f)}) we replace the derivatives of $g$, a long but straightforward calculation shows that the ideal sonic condition S$^{\rm G}$ is equivalent to
\begin{equation} \label{ISC}
\sum_{i=1}^8 R_i(r)  T_i(t) = 0 \, ,
\end{equation}
where the functions $R_i = R_i(r)$ are given by
%
%%%%%%%%%%%%%%%%%%%%%%%%%
%
\begin{widetext}
\begin{subequations}
\label{eq:R}
\begin{eqnarray}
R_1   \equiv  1 + \beta'(\alpha)   , \quad \qquad \ \   R_2  \equiv  \alpha + \beta \beta'(\alpha), \qquad \ \    R_3   \equiv  \beta + \alpha\beta'(\alpha), \qquad  \qquad   R_4  \equiv  \alpha \beta \, [1 + \beta'(\alpha)] , \qquad  \qquad \
\\[2mm] 
R_5   \equiv  \beta^2 + \alpha^2\beta'(\alpha)   , \qquad    R_6  \equiv  \beta^3 + \alpha^3\beta'(\alpha) , \qquad        R_7   \equiv  \alpha\beta \, [\beta + \alpha\beta'(\alpha)]   , 
\qquad    R_8  \equiv  \alpha\beta \, [\beta^2 + \alpha^2\beta'(\alpha)] , \qquad  \ \
\end{eqnarray}
\end{subequations}
and where the functions $T_i = T_i(t)$ are given by %
\begin{subequations}
\label{eq:T}
\begin{eqnarray}
T_1 & \equiv & -  g^2 (1 + g \dot{f})^2 \{\dot{f}^2 \ddot{f}^2 + 2 f\dot{f} \ddot{f}\, \dddot{f} + f(-2 \ddot{f}^3 - 3 f\dddot{f}^2  + 2 f\ddot{f}\, \ddddot{f})\}  ,\label{subeq:T1}     \\[2mm] 
T_2 & \equiv & f\{-g^3 \dot{f}^4 \ddot{f}^2 - g^2 \dot{f}^3 \ddot{f} (\ddot{f} + 2 g f\dddot{f}) + g \dot{f}^2 [\ddot{f}^2 + 2 g^2 f\ddot{f}^3  + 3 g^2 f^2 \dddot{f}^2 -  g f\ddot{f} (3 \dddot{f} + 2 g f\ddddot{f})] +  \nonumber  \\ 
    & + &  \dot{f} [\ddot{f}^2 + 2 g^2 f\ddot{f}^3  +   4 g^2 f^2 \dddot{f}^2 - g f\ddot{f} (2 \dddot{f} + 3 g f\ddddot{f})] + f[-\ddot{f}\, \dddot{f} + g^2 f\ddot{f}^2 \dddot{f} + g (\ddot{f}^3 + f\dddot{f}^2 - f\ddot{f}\, \ddddot{f})]\},      \label{subeq:T2}     \\[2mm]
T_3 & \equiv &  f\{-3 g^3 \dot{f}^4 \ddot{f}^2 - g^2 \dot{f}^3 \ddot{f} (5 \ddot{f} + 6 g f\dddot{f}) + 3 g \dot{f}^2 [-\ddot{f}^2 
		+ 2 g^2 f\ddot{f}^3 + 3 g^2 f^2 \dddot{f}^2 - g f\ddot{f} (3 \dddot{f} + 2 g f\ddddot{f})] +   
\nonumber  \\ 
    & + & \dot{f} [-\ddot{f}^2 + 10 g^2 f\ddot{f}^3 + 14 g^2 f^2 \dddot{f}^2 \!-\! g f\ddot{f} (2 \dddot{f} + 9 g f\ddddot{f})] + f[\ddot{f} \,\dddot{f} 
		\! - \! g^2 f\ddot{f}^2 \dddot{f} + g (3 \ddot{f}^3 \!+\! 5 f\dddot{f}^2 \! \!-\! 3 f\ddot{f}\, \ddddot{f})]\},  \quad   \label{subeq:T3}      \\[2mm] 
T_4 & \equiv &  f^{2}\{-3 g^2 \dot{f}^4 \ddot{f}^2 - 2 g \dot{f}^3 \ddot{f} (\ddot{f} + 3 g f\dddot{f}) + f(\ddot{f}^3 + 2 g f\ddot{f}^2 \dddot{f} 
		 + f\dddot{f}^2 - f\ddot{f}\, \ddddot{f}) + 
\nonumber  \\ 
    & + & \dot{f}^2 [\ddot{f}^2 + 6 g^2 f\ddot{f}^3 + 9 g^2 f^2 \dddot{f}^2 
		 - 6 g f\ddot{f} (\dddot{f} + g f\ddddot{f})] - 2 f\dot{f} [\ddot{f}\, \dddot{f} + g (-2 \ddot{f}^3 - 4 f\dddot{f}^2 
		+ 3 f\ddot{f}\, \ddddot{f})]\}  ,     \label{subeq:T4}      \\[2mm] 
T_5 & \equiv &  f^{2}\{-3 g^2 \dot{f}^4 \ddot{f}^2 - 2 g \dot{f}^3 \ddot{f} (2 \ddot{f} + 3 g f\dddot{f}) + 2 g f\dot{f} (4 \ddot{f}^3 + 5 f\dddot{f}^2 - 3 f\ddot{f}\, \ddddot{f}) +
\nonumber  \\ 
    & + &  f(\ddot{f}^3 - 2 g f\ddot{f}^2 \dddot{f} + 2 f\dddot{f}^2 - f\ddot{f}\, \ddddot{f}) +     \dot{f}^2 [-2 \ddot{f}^2 + 6 g^2 f\ddot{f}^3 + 9 g^2 f^2 \dddot{f}^2 - 6 g f\ddot{f} (\dddot{f} + g f\ddddot{f})]\}    ,     \label{subeq:T5}       \\[2mm] 
T_6 & \equiv & f^{3}\{-g \dot{f}^4 \ddot{f}^2 - f^2 \ddot{f}^2 \dddot{f} - \dot{f}^3 \ddot{f} (\ddot{f} + 2 g f\dddot{f}) + f\dot{f} (2 \ddot{f}^3 + 2 f\dddot{f}^2 -
		 f\ddot{f}\, \ddddot{f}) +
\nonumber  \\ 
    & + &  f\dot{f}^2 [-\ddot{f}\,  \dddot{f} + g (2 \ddot{f}^3 + 3 f\dddot{f}^2 - 2 f\ddot{f}\, \ddddot{f})]\}  ,     \label{subeq:T6}      \\[2mm] 
T_7 & \equiv &  f^{3}\{-3 g \dot{f}^4 \ddot{f}^2 + f^2 \ddot{f}^2 \dddot{f} - \dot{f}^3 \ddot{f} (\ddot{f} + 6 g f\dddot{f}) + f\dot{f} (2 \ddot{f}^3 + 4 f\dddot{f}^2 	- 3 f\ddot{f}\, \ddddot{f}) + 
\nonumber  \\ 
    & + &  3 f\dot{f}^2 [-\ddot{f}\, \dddot{f} + g (2 \ddot{f}^3 + 3 f\dddot{f}^2 - 2 f\ddot{f}\, \ddddot{f})]\}   ,     \label{subeq:T7}      \\[2mm] 
T_8 & \equiv &  - f^{4} \dot{f}^2 \{\dot{f}^2 \ddot{f}^2 + 2 f\dot{f} \ddot{f}\, \dddot{f} + f(-2 \ddot{f}^3 - 3 f\dddot{f}^2 + 2 f\ddot{f} \, \ddddot{f})\}    .     \label{subeq:T8}              
\end{eqnarray}
\end{subequations}
\end{widetext}

\subsection{Analyzing the ideal model equation} 
\label{subsec-idealvaris}

The above analysis of the ideal sonic condition S$^G$ leads to Eq. (\ref{ISC}). This constraint and (\ref{gt(f)}) constitute a differential system for the functions $\{f(t), g(t); \beta(\alpha)\}$. The study of the general solution to this system is a complex task that requires the use of numerical methods and that falls outside the scope of this work. 

Alternatively, we can use an analytical approach in looking for some particular solutions. For example, we can choose a particular function $\beta= \beta(\alpha)$, then determine the functions $R_i(r)$, and finally solve the system of equations that (\ref{ISC}) and (\ref{gt(f)}) impose on $\{f(t), g(t)\}$ (see Sec. \ref{subsec-ISC}). 

We can also consider a family to solutions of the field equations, and then analyze whether a subfamily fulfills the ideal sonic condition (\ref{ISC}). This is the approach that we follow in the next section for the spherically symmetric limit of the Szafron solution \cite{Szafron}.

%%%%%%%%%%%%%%%%%%%%%%%%%

\section{The ideal Szafron model $f(t)=t^q$} 
\label{sec-t^q}

Now we shall study whether the Szafron solution \cite{Szafron}, which is defined by the choice $f(t)=t^q$, $q \not=1/2$, is compatible with the ideal sonic condition (\ref{ISC}). From (\ref{g(f)}) we obtain that the general solution of the field equations takes form (\ref{sol. general eq. Z}) with
\begin{equation} \label{g-ideal}
f(t) = t^q  , \quad g(t) = - \sigma^{-1} t^{1-q}, \quad  \sigma \equiv 2q\!-\!1 \not=0  \, .
\end{equation}
From these expressions we obtain that the functions $T_i =T_i(t)$ given in (\ref{eq:T}) become:
\begin{subequations}
\label{eq:T-Szafron}
\begin{eqnarray}
T_1 = T_2 = T_3=T_6 = T_7 = T_8 = 0 , \qquad \\[1mm]  T_4 = -T_5 = \frac18 \sigma (1-\sigma^2)^2 t^{-6} . \qquad
\end{eqnarray}
\end{subequations}
Thus, we have that all the functions $T_i$ identically vanish if, and only if, $\sigma= \pm 1$, which corresponds to the dust LT-model with $\Lambda =0$. Otherwise, the ideal sonic condition (\ref{ISC}) holds when $R_4 = R_5$.  Expressions in (\ref{eq:R}) for these functions imply that the functions $\alpha(r) \equiv a(r)/b(r)$ and $\beta(r) \equiv a'(r)/b'(r)$ fulfill the following relation:
\begin{equation} \label{beta(alpha)-ideal}
\beta(\alpha) =  c \alpha \, ,  
%\quad c= \text{constant} .
\end{equation}
where $c\not=1$ is a constant (note that $c=1$ leads to the FLRW limit). This equation also holds if we change the functions $g(t)$, $\alpha$, and $\beta$ by a factor; thus, we can take $g(t) = t^{1-q}$. 

Then, taking into account that (\ref{beta(alpha)-ideal}) is a differential equation that relates $a(r)$ and $b(r)$, we obtain a solution of the perfect fluid Einstein equations, which is compatible with the ideal sonic condition {\rm S}$^\textrm{G}$ (\ref{chi-gas-ideal}), given by the metric {\rm (\ref{metric-Y})} with the following election of the metric function $Y(t,r)$,
\begin{equation} \label{Y-ideal}
Y=Z^{2/3} , \quad      Z(t,r) = t^{\frac{1-\sigma}{2}} b(r) [1 + \alpha(r) t^{\sigma}]  , 
\end{equation}
where $b(r)$ is given by
\begin{equation} \label{b-ideal}
b(r) = |\alpha(r)|^{1/(c-1)} \,  ,  \quad c\not=1 \,  .
\end{equation}
%
%Moreover, we have $g_{rr}= Y'^2$, with
%
%\be
%Y' = -  \frac{2\,t^{\frac{1-\sigma}{2}} b(r) }{3(1-c)} \frac{ \alpha'(r)[1 + c \alpha(r) t^{\sigma}]}{ \alpha Z^{1/3}} \, .
%\ee
%
 
Regarding the expansion of the fluid flow, from (\ref{expansion-Zb}) we have 
\begin{equation} \label{expansion-ideal}
\theta = \frac{1}{t}\left[1\!+\!\sigma - \frac{\sigma}{1 + \alpha t^{\sigma}} - \frac{\sigma}{1 + c \alpha t^{\sigma}}\right]  \, . 
\end{equation} 

It is worth remarking that we are considering, as Szafron did, expanding models with $t > 0$. Nevertheless, the change $t \leftrightarrow -t$, with $t<0$, leads to contracting models, and our analysis, given below, of the physical properties of the solutions is also valid in this case.

%%%%%%%%%%%%%%%%%%%%%

\begin{figure*}
\centerline{
\parbox[c]{0.33\textwidth}{\includegraphics[width=0.30\textwidth]{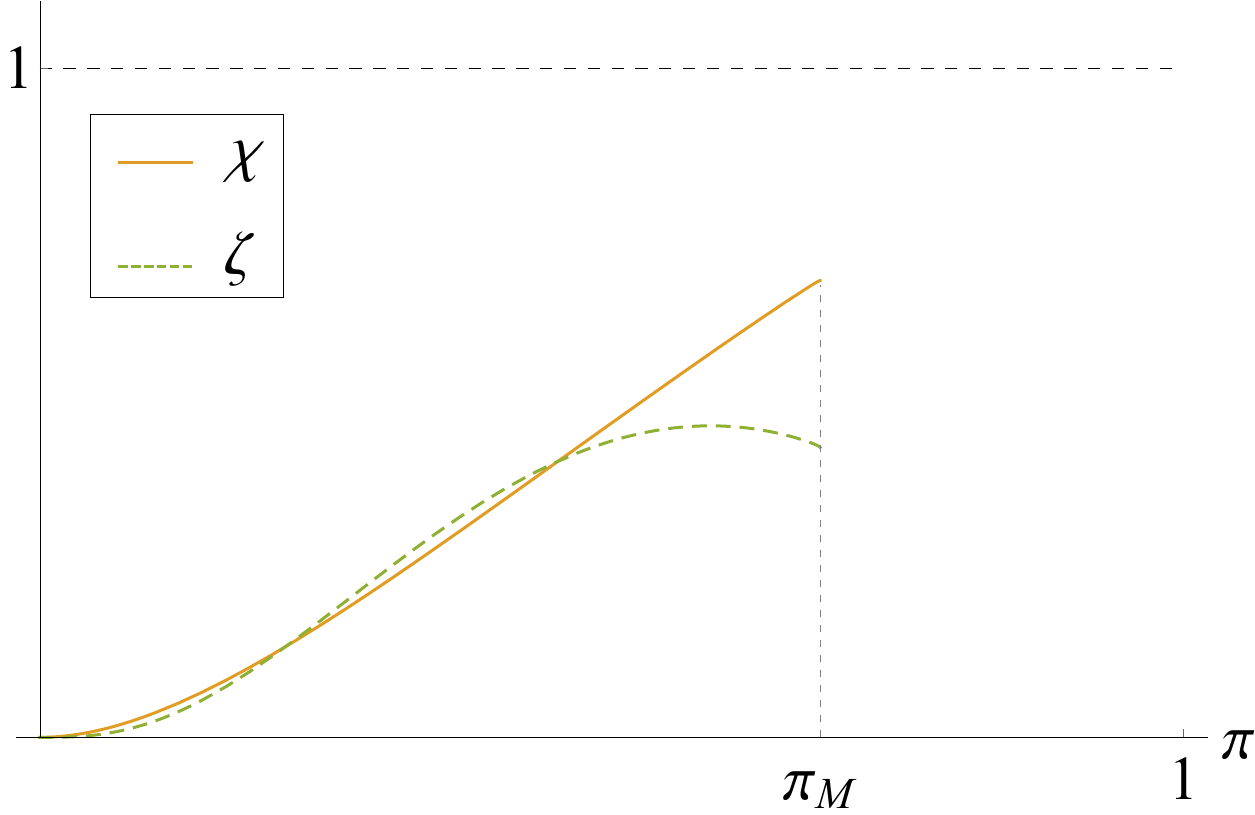}\\(a)}
\parbox[c]{0.33\textwidth}{\includegraphics[width=0.30\textwidth]{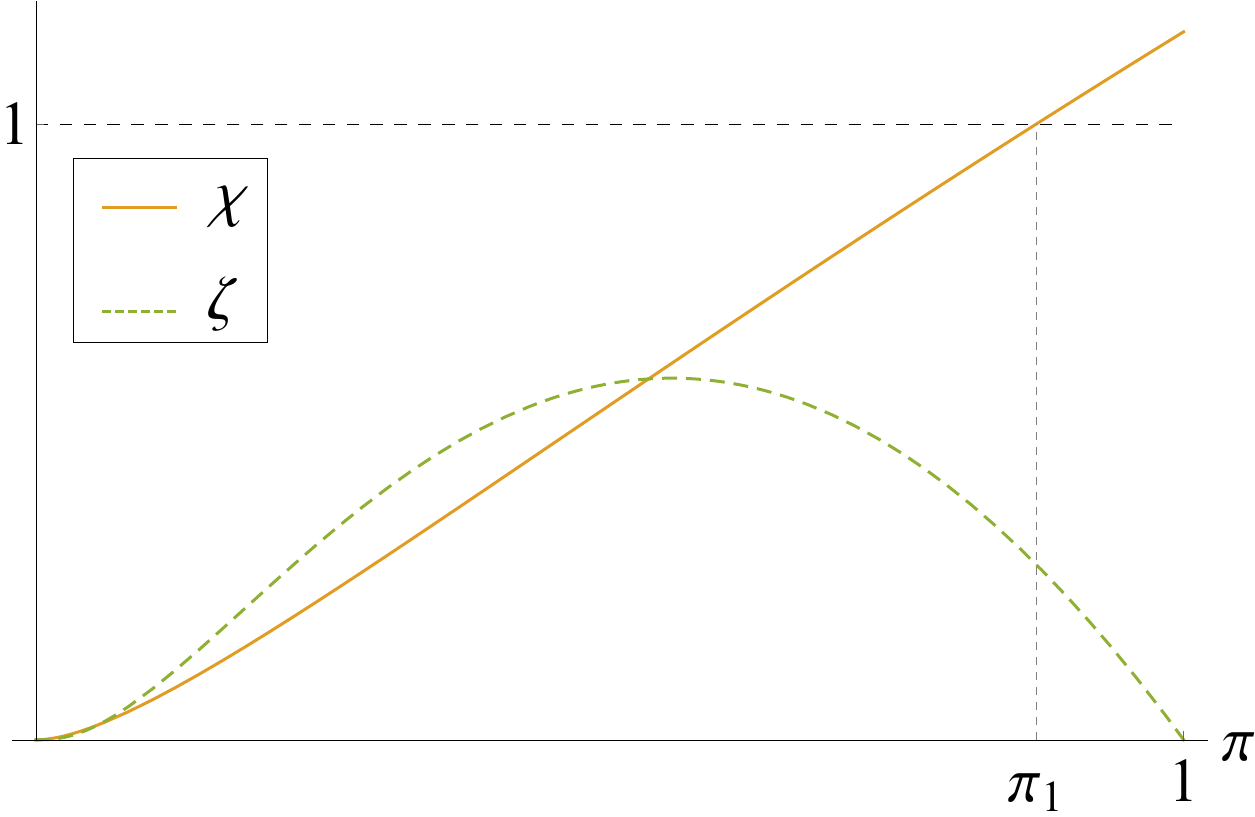}\\(b)}
\parbox[c]{0.33\textwidth}{\includegraphics[width=0.30\textwidth]{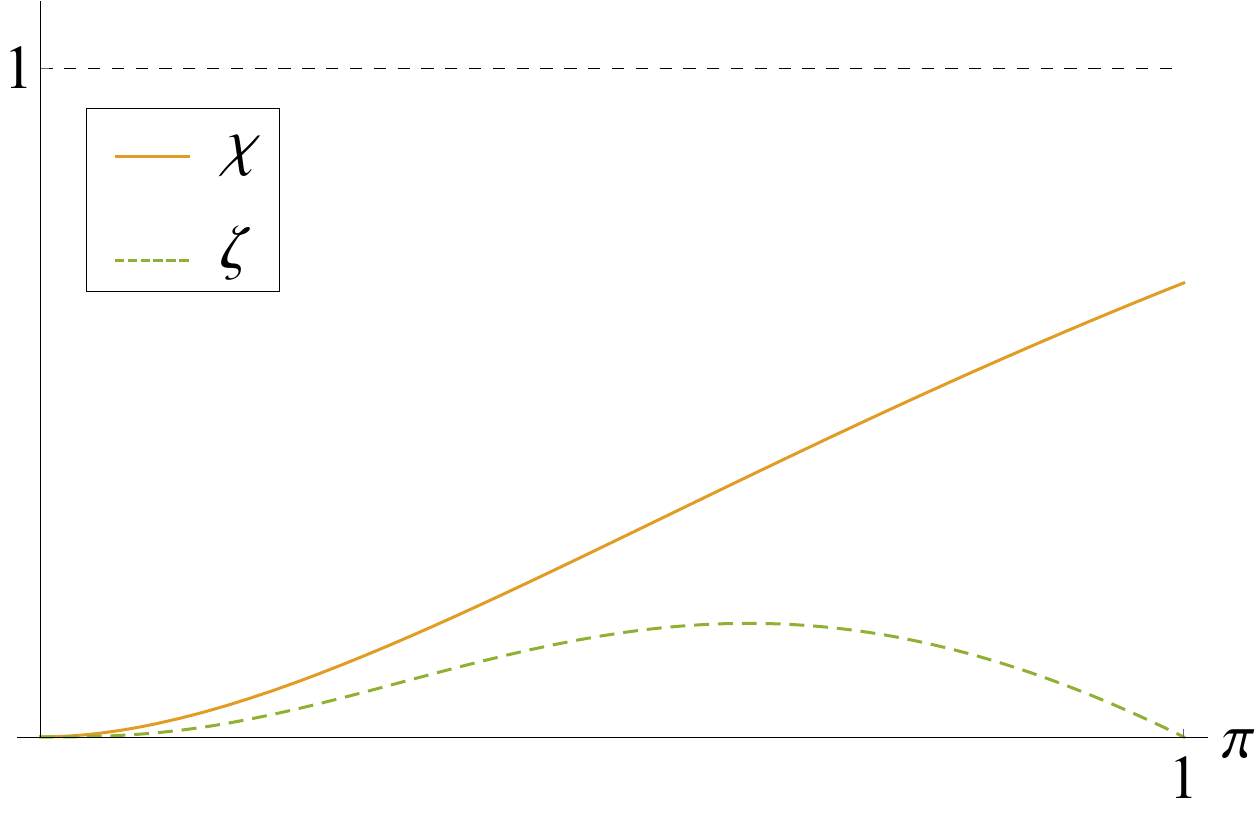}\\(c)}}
\caption{The orange solid line shows the behavior of the indicatrix function $\chi=\chi(\pi)$ of the ideal Szafron models. (a) Case $-1 \leq c \leq 0$; (b) Case $c>0$, $\epsilon = +1$; (c) Case $c>0$, $\epsilon = -1$, $|\sigma| > \sigma_0$. We have also plotted function $\zeta(\pi)$ given in (\ref{cc-ideal}) (green dashed line), which is positive in the interval where $\chi(\pi)\in ]0,1[$.}
\label{fig-1}
\end{figure*}

%%%%%%%%%%%

%%%%%%%%%%%%%%%%%%%%%%%%%

\subsection{Hydrodynamic quantities: Energy density, pressure, and speed of sound} 
\label{subsec-hydro-ideal}

Now we can obtain expressions for the hydrodynamic quantities $\rho$, $p$, and  the indicatrix function $c_s^2 = \chi(\pi)$. The time dependence of the pressure and the energy density can easily be obtained from the expressions in (\ref{pressure and density-Zb}). The pressure is
\begin{equation} \label{p-ideal}
p =  \frac{(1- \sigma^2)}{3 t^2}   \, ,
\end{equation}
and the energy density is
\begin{equation} \label{rho-ideal-1}
 \rho = 
%\rho(t,\alpha) \equiv 
 \frac{1}{3t^2}\frac{[(1\!-\!\sigma) \! +\!(1\!+\!\sigma) \alpha t^{\sigma} ]  [(1\!-\!\sigma) \! + \!(1\!+\!\sigma) c \alpha t^{\sigma}]}{(1 + \alpha t^{\sigma})(1 + c \alpha t^{\sigma})}  . 
\end{equation}

Now, solving the equation $\rho = \rho(t,\alpha)$ for $\alpha$ and using (\ref{p-ideal}) to eliminate $t$, the following function of state can be obtained when $c \not=0$,
\begin{eqnarray} \label{alpha(rho,p)-1}
\alpha = \alpha (\rho,p) \equiv
\kappa_0 \sqrt{(3p)^{\sigma}} \frac{(c\! +\! 1)(1 - \pi) +  \varepsilon \sigma F(\pi)}{ (1\!-\!\sigma) -(1\!+\!\sigma) \pi}  ,  \qquad \\[2mm]
\label{F(pi)-1}
	 F(\pi) = \sqrt{\hat{c}^2 (1 - \pi)^2 +16 c \pi/(1\! -\!\sigma^2)}    \, ,  \qquad \\[2mm]
%
%\begin{equation} 
\label{F(pi)-1-kappa}
	\kappa_0 \equiv -(1-\sigma^2)^{1-\frac{\sigma}{2}}/[2 (1+\sigma) c]  \, , \qquad  \\[2mm]
	\varepsilon = \pm 1, \qquad \hat{c} = (1-c)/\sigma \, .  \qquad 
\end{eqnarray}

Finally, from expression (\ref{chi-general}) with (\ref{g-ideal}), (\ref{beta(alpha)-ideal}), and (\ref{expansion-ideal}), and taking into account (\ref{p-ideal}) and (\ref{alpha(rho,p)-1}) to eliminate $t$ and $\alpha$, the indicatrix function $\chi(\pi)$ can be determined,
\begin{equation}   \label{chi(pi)-1}
\hspace{0mm}\ c_s^2 = \chi(\pi) \!\equiv\! \frac{4\pi^2[\hat{c}^2 (1\!+\!\pi) + (1\!+\!c) \varepsilon F(\pi)]}{(1\!+\!\pi) [\hat{c}^2(1\!-\!\sigma^2) (1\!+\!\pi)^2 \!+\! 4(1\! +\! c)^2 \pi]} , 
\end{equation}
When $c=0$ the above expression for $\chi(\pi)$ remains valid by taking $\varepsilon = +1$.

Here we are interested in non barotropic ($\alpha \not=$ cons\-tant) solutions with a non-vanishing pressure ($\sigma^2 \not=1$). Besides, only a positive pressure ($\sigma^2 < 1$) is compatible with the ideal gas equation of state (\ref{gas-ideal}). On the other hand, the changes $(\sigma, c, \alpha) \leftrightarrow (-\sigma, c^{-1}, \alpha^{-1})$ leave the metric unchanged. Thus, $\alpha(r)$ being a non-constant arbitrary function, we can analyze all the ideal Szafron models by considering $\sigma^2 <1$ and $-1 \leq c <1$.

Let us note that, as Szafron already pointed out \cite{Szafron}, the solutions approach a FLRW model with a $\gamma$-law, $p=(\gamma-1) \rho$, when $t \rightarrow 0$ or $t \rightarrow \infty$. Indeed, from (\ref{p-ideal}) and (\ref{rho-ideal-1}), we obtain:
\be \label{limits}
\rho (t \rightarrow 0) = \frac{1\!-\!|\sigma|}{1\!+\!|\sigma|}\, p , \qquad \rho (t \rightarrow \infty) = \frac{1\!+\!|\sigma|}{1\!-\!|\sigma|}\, p .
\ee
\vspace*{-6mm}

%%%%%%%%%%%%%%%%%%%%%%

\subsection{Fluid properties: Compressibility conditions {\rm H}$_1^{\rm G}$}
\label{subsec-compress-ideal}

Expression (\ref{chi(pi)-1}) of the indicatrix function $\chi(\pi)$ defines a function of state that characterizes a family of fluids. We can analyze the physical reality of these fluids regardless the particular evolution that the ideal Szafron models represent. More specifically, now we study the compressibility conditions H$_1^{\rm G}$ given in (\ref{cc-ideal}). 

We must analyze the behavior of the function $\chi(\pi)$ in the interval $0 < \pi < 1$ where the energy conditions E$^\textrm{G}$ hold. Note that $\chi(\pi)$ depends on the parameters $c$ and $ \sigma^2$, and the sign $\varepsilon$.

Firstly, we analyze the first constraint in (\ref{cc-ideal}), the causal condition $0<\chi(\pi) <1$. When $-1 \leq c \leq 0$, $\chi(\pi)$ is an increasing function and it is well defined and fulfills this compressibility condition in an interval $]0, \pi_M[$, where $\pi_M \equiv 1\!+ \! w \!-\! \sqrt{w (w\!+\!2)}$, with $w \equiv -8 c /[\hat{c}^2 (1\!-\!\sigma^2)]>0$ (see Fig. \ref{fig-1}a). Note that $\pi_M$ is close to $1$ (respectively, is close to $0$) when $c$ or $\sigma$ are close to zero (respectively, $\sigma^2$ is close to $1$). 

When $1> c > 0$, the behavior of the indicatrix function depends on $\varepsilon$. If $\varepsilon =+1$, then $\chi(\pi)$ is a positive increasing function in the whole interval $]0, 1[$, and fulfills the causal constraint in the interval $\,0 < \pi< \pi_1 <1$, with $\chi(\pi_1)=1$ (see Fig. \ref{fig-1}b). Moreover, $\pi_1$ is close to $1$ (respectively, to $0$) when $c$ or $\sigma$ are close to zero (respectively, $c$ is close to $1$) . 

If $c>0$ and $\varepsilon =-1$, we have two possibilities; when $|\sigma| < \sigma_0 = (1-c)/(1+c)$, then $\chi(\pi)$ is a positive increasing function and fulfills the causal constraint in the whole interval $]0, 1[$ (see Fig. \ref{fig-1}c); and when $|\sigma| > \sigma_0$, then $\chi(\pi)$ is a negative function and thus it does not meet the causal constraint at any point. 

Regarding the second of the compressibility conditions H$_1^{\rm G}$, $\zeta(\pi)>0$, it holds throughout the interval where $\chi(\pi)$ is well defined and fulfills the causal condition in each of the cases considered above (see Fig. \ref{fig-1}).

It is worth remarking that the ideal Szafron models do not represent the evolution of a classical ideal gas because the equation of state (\ref{chi(pi)-1}) is not compatible with the one of a classical ideal gas, namely, $\chi(\pi) = \gamma \pi/(1+\pi)$ \cite{CFS-CIG}. This result agrees with a result on the study of the velocities of the classical ideal gases \cite{FS-KCIG}; a geodesic and expanding timelike unit vector is the unit velocity of a classical ideal gas if, and only if, it is vorticity-free and its expansion is homogeneous. 

Note that the function $\chi(\pi)$ (\ref{chi(pi)-1}) verifies $\chi(0)=\chi'(0)=0$. Thus, it approaches that of a classical ideal gas (or a monoatomic Synge gas) at zero-order (but not at first-order) for small values of $\pi$. For every value of $\sigma$, a value of $c$ exists for which the indicatrix function (\ref{chi(pi)-1}) approaches that of the Synge gas at zero-order in the ultrarelativistic regime, $\chi(1/3) = 1/3$.

%%%%%%%%%%%%%%%%%%%%%%

%%%%%%%%%%%%%%%%%%%%%

\begin{figure*}
\centerline{
\parbox[c]{0.33\textwidth}{\includegraphics[width=0.30\textwidth]{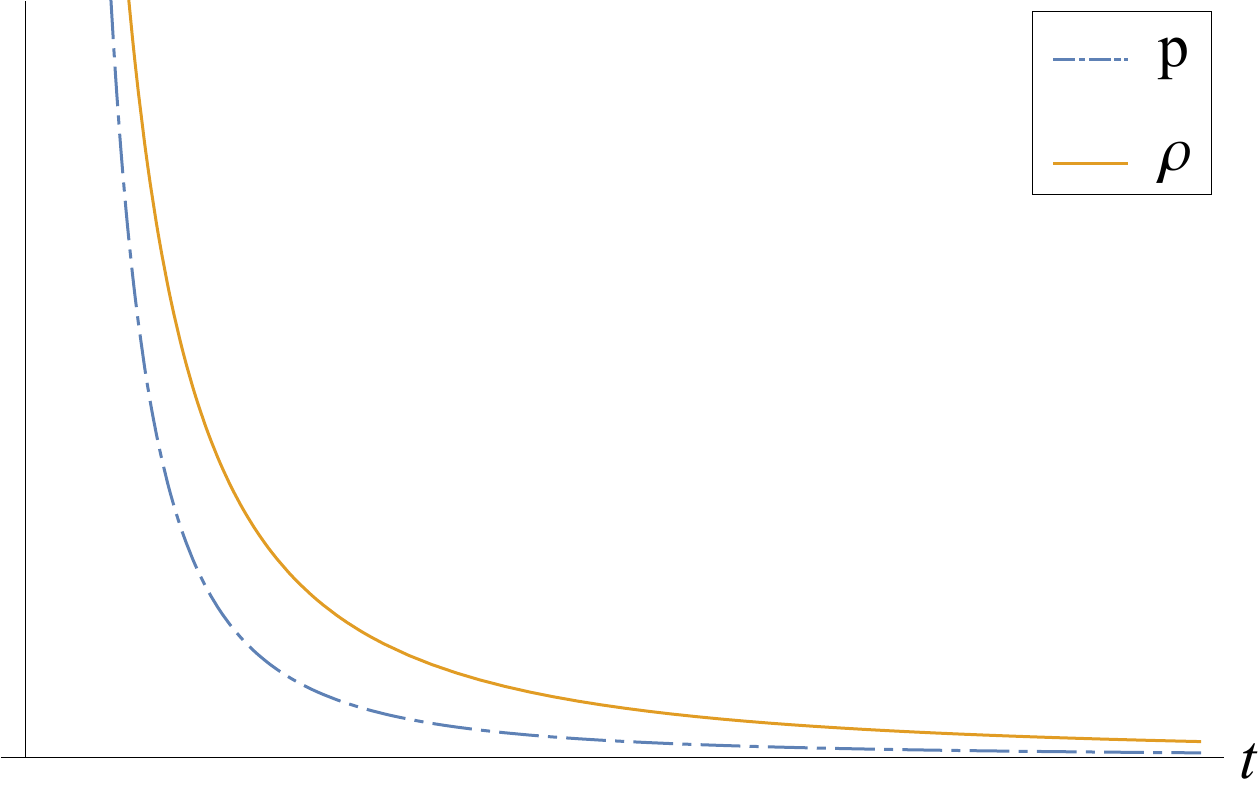}} 
\parbox[c]{0.33\textwidth}{\includegraphics[width=0.30\textwidth]{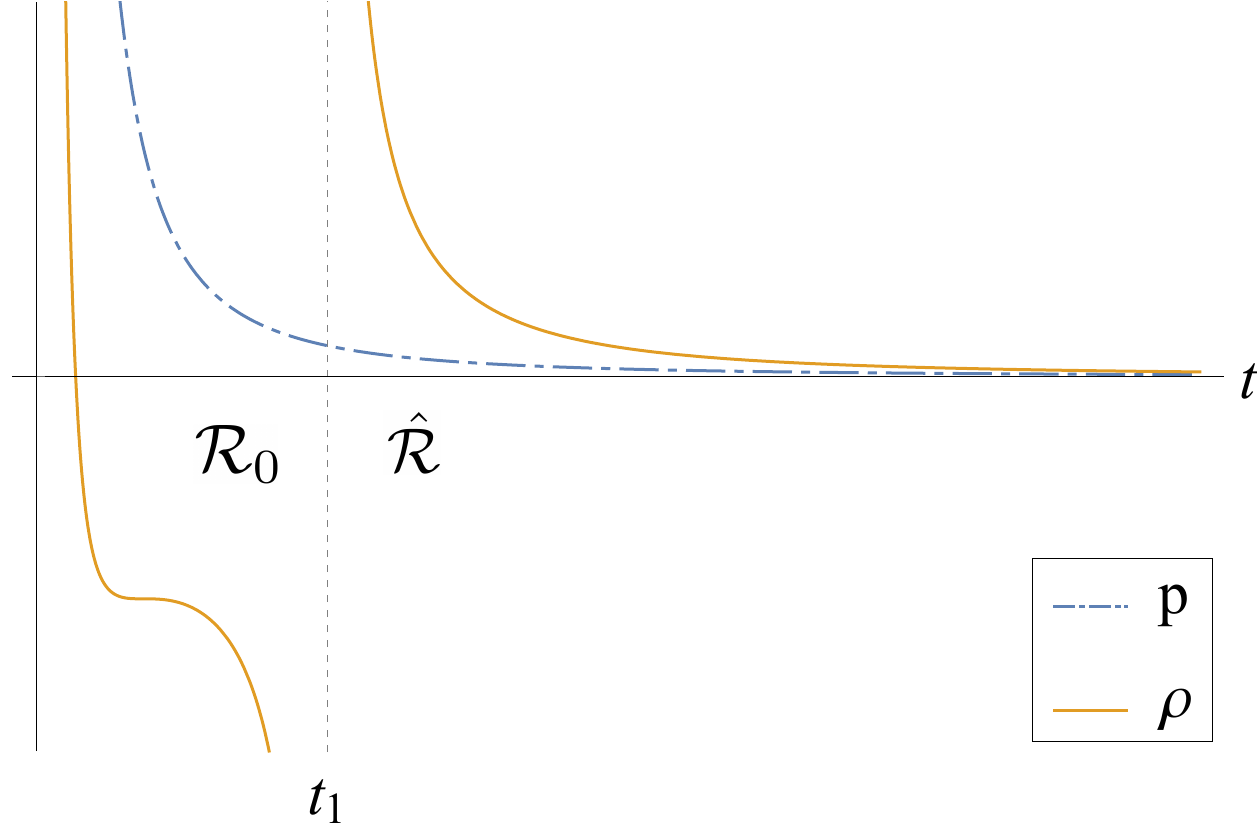}}
\parbox[c]{0.33\textwidth}{\includegraphics[width=0.30\textwidth]{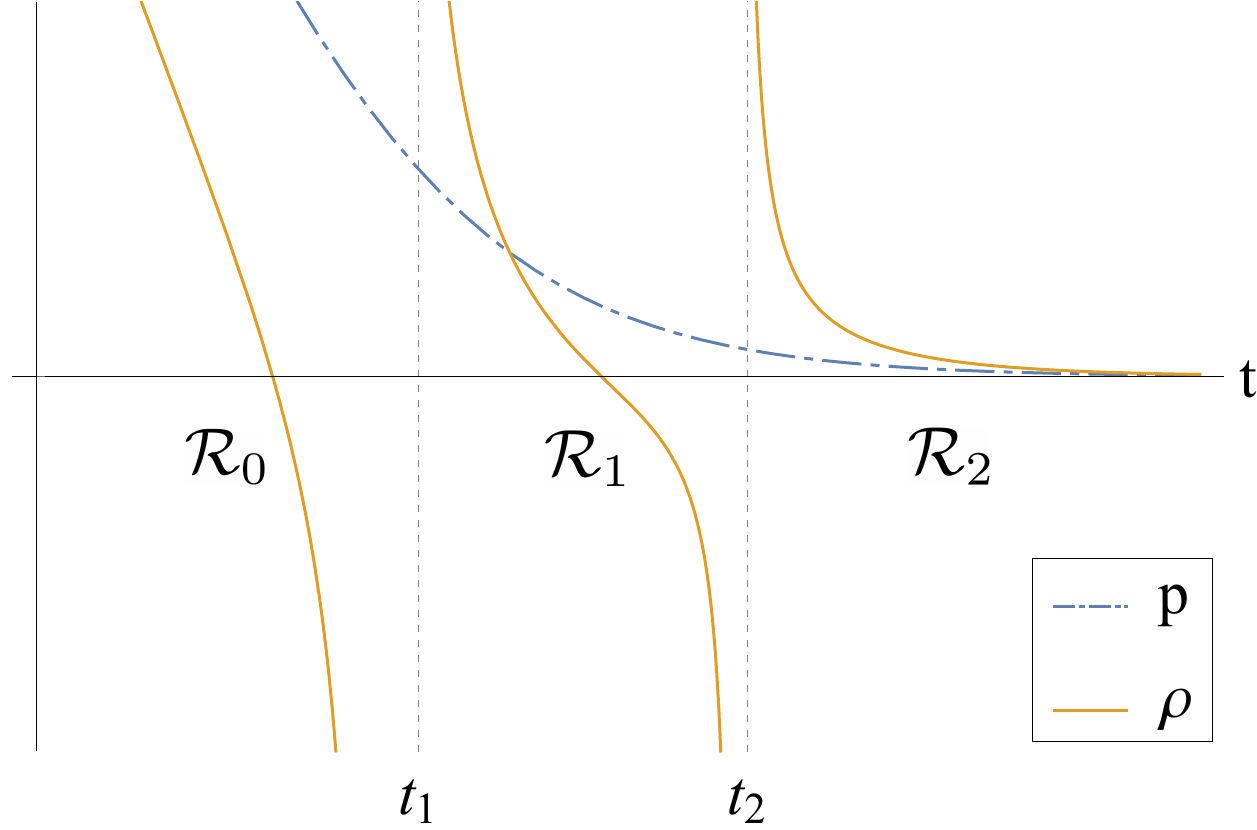}}}
%\end{figure*}
%\begin{figure*}
\centerline{
\parbox[c]{0.33\textwidth}{\includegraphics[width=0.30\textwidth]{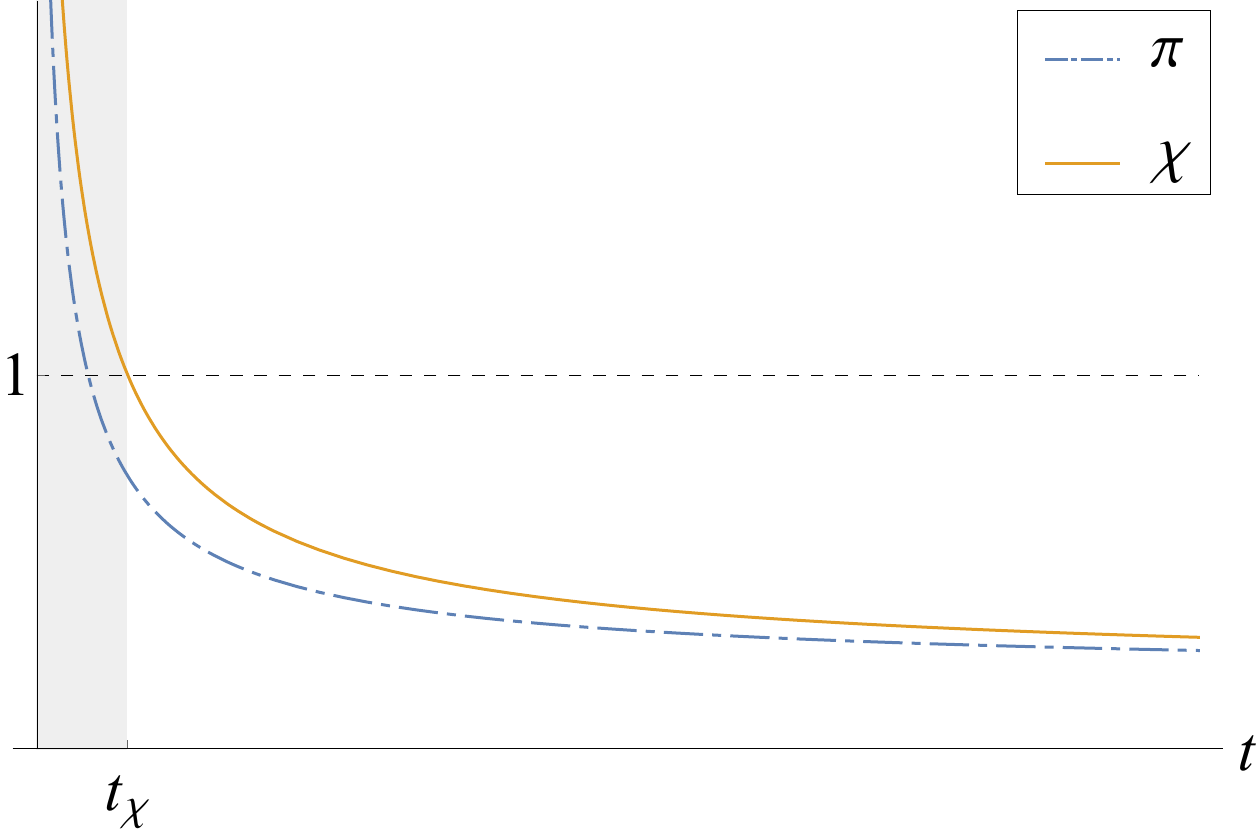}\\(a)}
\parbox[c]{0.33\textwidth}{\includegraphics[width=0.30\textwidth]{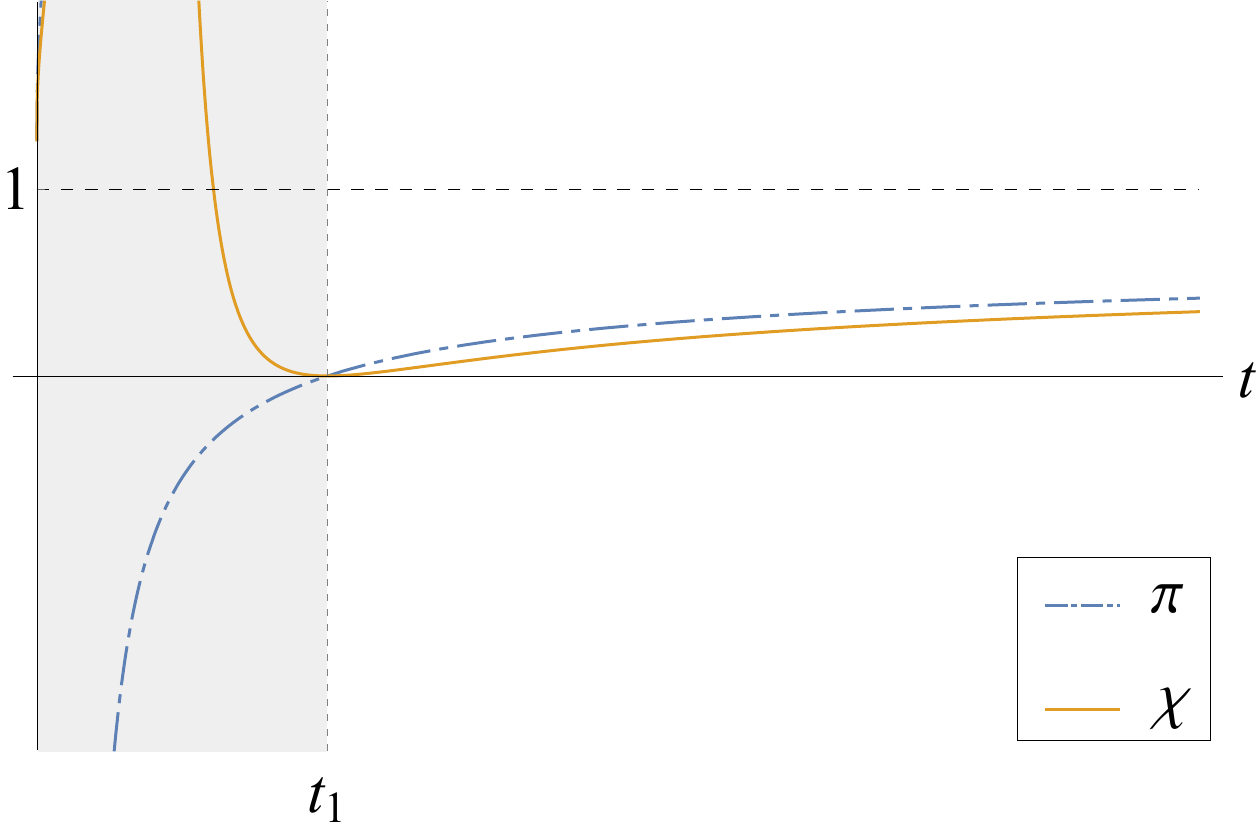}\\(b)}
\parbox[c]{0.33\textwidth}{\includegraphics[width=0.30\textwidth]{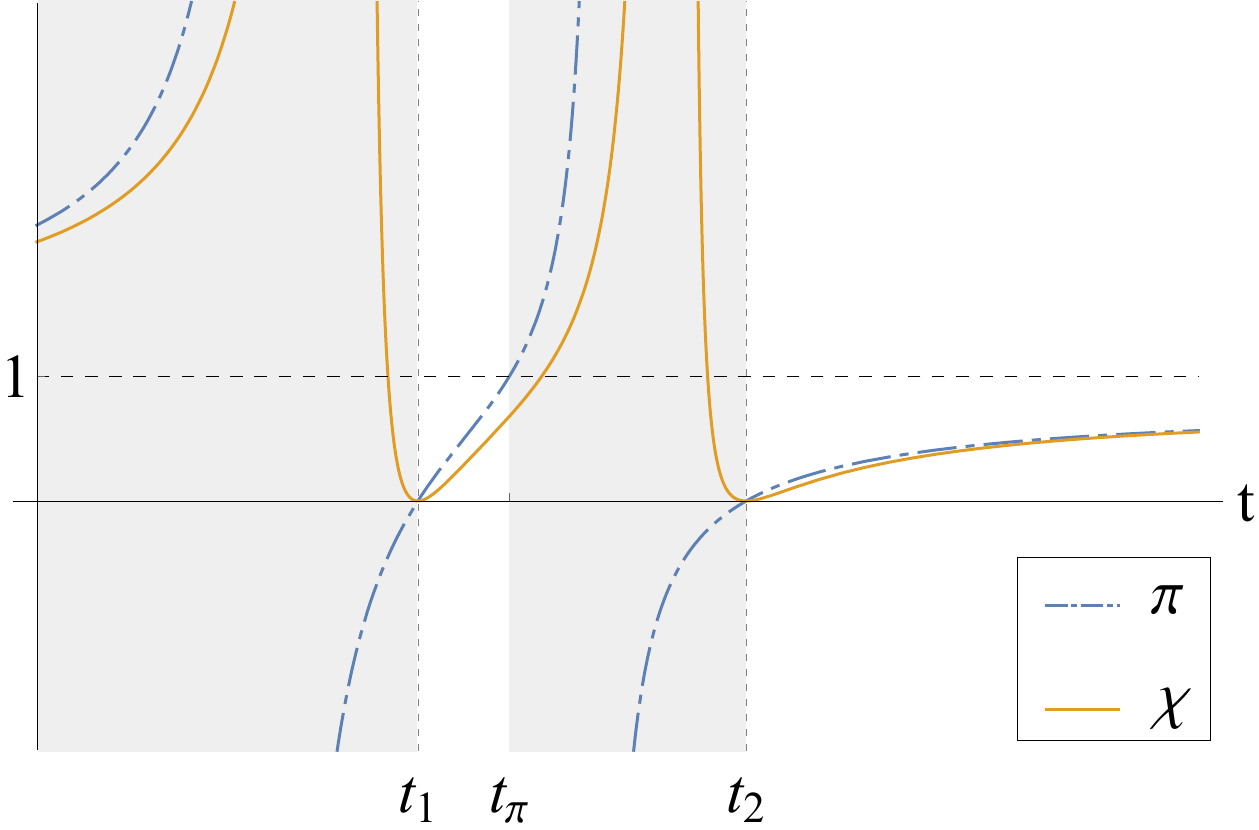}\\(c)}}
\caption{Time evolution of the hydrodynamic variables of the ideal Szafron models for a fixed $r$. At the top we have plotted the energy density $\rho(t,r)$ (orange solid line) and the pressure $p(t,r)$ (blue dot-dashed line); these figures also show the spacetime regions defined by the spacetime singularities. At the bottom we have plotted the hydrodynamic variables $\chi(t,r)$ (orange solid line) and $\pi(t,r)$ (blue dot-dashed line), and we have shaded the spacetime domains where the conditions for physical reality do not hold. (a) Case $c \geq 0$, $\alpha > 0$. (b) Case $c\leq0$, $\alpha<0$ (the case $c<0$, $\alpha>0$ is similar exchanging $t_1$ for $t_2$).  (c) Case $c>0$, $\alpha<0$, $\sigma>0$ and $t_1 < \hat{t}_{\rho}$; the temporal axis is represented in logarithmic scale (the case $c>0$, $\alpha<0$, $\sigma>0$, and $t_1 > \hat{t}_{\rho}$ is similar but without the non-shaded domain $]t_1, t_{\pi}[$); for this case (c), but with $\sigma < 0$, we have a similar situation, but exchanging $t_\rho \, \leftrightarrow \, \hat{t}_\rho$ and $t_1 \, \leftrightarrow \, t_2$.}
\label{fig-2}
\end{figure*}

%%%%%%%%%%%

\subsection{Curvature singularities and spacetime domains} 
\label{subsec-singularities}

The expressions for the metric line element, expansion, energy density and pressure of the ideal Szafron models given at the beginning of this section show that our models can have up to three different singularities. The first one takes place at $t = 0$. At this singularity the line element of the 3-spaces $t = constant$ vanishes and the energy density and pressure become infinite. Thus, this is a big bang singularity. 

Secondly, when $t_1^{\sigma}\, \alpha(r)= -1$, the metric line element of the sphere vanishes while the metric distance on the coordinate lines of coordinate $r$ becomes infinite. Moreover, at this singularity, which is not simultaneous for the comoving observer ($t_1 = t_1(r)$), the energy density is infinite. 

Finally, a singularity appears when $t_2^{\sigma} \, \alpha(r) = -c^{-1}$. At this singularity, which is not simultaneous either ($t_2 = t_2(r)$), the metric distance on the coordinate lines of coordinate $r$ vanishes and we have infinite energy density again.

Depending on how many of these singularities appear, three different cases can be distinguished (see Fig. \ref{fig-2}): 
\begin{itemize}
\item[(i)] If $c \geq 0$ and $\alpha(r) > 0$, then only the singularity at $t = 0$ occurs. Consequently, the solution is defined in the full domain $t>0$ (see Fig. \ref{fig-2}a).
\item[(ii)] If $c \leq 0$ and $\alpha < 0$, we have a singularity at $\hat{t}(r) = t_1$, and if $c < 0$ and $\alpha > 0$ we have a singularity at $\hat{t}(r) = t_2$. In both cases the singularity at $t = 0$ occurs. Now, two disconnected spacetimes domains exist: ${\cal R}_0 = \{0 < t < \hat{t}\}$ and $\hat{\cal R} = \{\hat{t}<  t < \infty\}$ (see Fig. \ref{fig-2}b).
\item[(iii)] Finally, if $c > 0$ and $\alpha(r) < 0$, then we have all three singularities, and three disconnected spacetimes domains exist: ${\cal R}_0 = \{0 < t < t_1\}$, ${\cal R}_1 = \{t_1 < t < t_2\}$  and ${\cal R}_2 = \{t_2<  t < \infty\}$ (see Fig. \ref{fig-2}c).
\end{itemize}
%

%%%%%%%%%%%%%%%%%%%%%%

\subsection{Analysis of the evolution: Energy and compressibility conditions} 
\label{subsec-energy-c}

Now we study the spacetime domains where the energy conditions E$^G$ given in (\ref{e-c-G}) and the compressibility conditions H$_1^G$ given in (\ref{cc-ideal}) hold. These domains are related to those considered above and defined by the spacetime singularities, but they also depend on the times at which the energy density vanishes or at which $\pi(t,r)=1$. All these times are not simultaneous for the comoving observer and depend on $\alpha(r)$. The last one, $t_{\pi}(r)$, is defined by the constraint $c\, t_{\pi}^{\sigma} \alpha(r) = \sqrt{(1\!-\!\sigma)/(1\!+\!\sigma)}$, which can be obtained from the expression $\pi(t, \alpha) = p(t)/ \rho(t,\alpha)$, where $p(t)$ and $\rho(t,\alpha)$ are given in (\ref{p-ideal}) and (\ref{rho-ideal-1}), respectively. On the other hand, the energy density vanishes at two times, $t_{\rho}(r)$ and $\hat{t}_{\rho}(r)$, on the region ${\cal R}_0$ when $c>0$ and $\alpha <0$, and it vanishes at one of them when $c<0$, or when $c=0$ and $\alpha < 0$. These times are defined by the constraints $\, t_{\rho}^{\sigma} \alpha(r) = (1\!-\!\sigma)/(1\!+\!\sigma)$ and $c\, {\hat{t}}_{\rho}^{\sigma} \alpha(r) = (1\!-\!\sigma)/(1\!+\!\sigma)$. Finally, we must also consider the times $t_{\chi}(r)$, $\bar{t}_{\chi}(r)$ and $\hat{t}_{\chi}(r)$, defined by the condition $\chi(t, r) = 1$.  

The role played by the above times in defining the spacetime regions where the energy conditions E$^{\rm G}$ and the compressibility conditions H$_1^{\rm G}$ hold depend on the signs of $\sigma$  and $c$. In Table \ref{table-1} the results for $\sigma>0$ are summarized, and in Fig. \ref{fig-2} we have plotted the different possible cases.

%%%%%%%%%%%%%%%%%%%%%

%%%%%%%%%%%%%%%%
%
\begin{table*}[t]
\caption{This table provides the space-time regions where the hydrodynamic conditions for physical reality hold for those models with $\sigma > 0$, which differ depending on the sign of the parameter $c$. The boundary times $t_\pi(r)$, $t_\rho(r)$, $\hat{t}_\rho(r)$, $t_1(r)$, $t_2(r)$, $t_\chi(r)$, $\bar{t}_{\chi}(r)$, and $\hat{t}_{\chi}(r)$, are defined in Secs. \ref{subsec-singularities} and \ref{subsec-energy-c}. For those models with $\sigma <0$ the results are the same, but exchanging $t_\rho \, \leftrightarrow \, \hat{t}_\rho$ and $t_1 \, \leftrightarrow \, t_2$ in the $c > 0$ and $\alpha(r) < 0$ case.}
\vspace{2mm}
\label{table-1}
\begin{tabular}{ccccccl}
 \hline \\[-3.8mm] \hline \\[-3mm]
  & & $\rho > 0$ & $0 < \pi < 1$ & $0 < \chi < 1$ & 
  \hspace{-10mm}  
  \\[2mm]
 \hline \\[-3mm]
   \  $c \geq 0 ,\ \ \alpha(r) > 0 \quad $ & & $[0, \infty[$  & $]t_\pi, \infty[$  & $]t_\chi,  \infty[ $ & 
\\[0.8mm] 
 %
%\hspace{-3mm} \quad $c \geq 0$ \quad & & & & & & 
%   
\hspace{-20mm}    
 \hspace{3cm}\\[-3.5mm]
\hspace{-1mm}      &  \ \ $t_1 > \hat{t}_\rho$ & \qquad $[0, t_\rho[ \, \cup \, ]\hat{t}_\rho, t_1[ \, \cup \, ]t_2, \infty[ \quad$ & \quad $]t_\pi, t_1[ \, \cup \, ]t_2, \infty[ \quad$  & $]t_\chi,  \infty[ $ & 
\hspace{-20mm}    
 \hspace{3cm}\\[-1mm]
 %\hline
 %
\    $c > 0, \ \ \alpha(r) < 0 \quad $ &  &  &  & & 
\hspace{-20mm}   
 \hspace{3cm}\\[-2mm]
 %\hline
 %
\hspace{-1mm}    & \ \  $t_1 < \hat{t}_\rho$ & \qquad $[0, t_\rho[ \, \cup \, ]t_1, \hat{t}_\rho[ \, \cup \, ]t_2, \infty[ \quad$ & \quad $]t_\pi, \hat{t}_\rho[ \, \cup \, ]t_2, \infty[ \quad$ & \quad $]\hat{t}_\chi,  \bar{t}_\chi[ \, \cup \, ]t_\chi, \infty[ \quad$ &   
\hspace{-20mm}   
\\[1mm] \hline \\[-3mm]
\ $c < 0, \ \ \alpha(r) > 0 \quad$  & & $[0, \hat{t}_\rho[ \, \cup \, ]t_2, \infty[$ & $]t_2, \infty[$ & $]t_\chi, \infty[$ &  \\[2mm]
%   
%\hspace{-3mm} \quad $c < 0$ \quad  & & & & & & \\[-2mm]
%   
\ $c \leq 0, \ \ \alpha(r) < 0 \quad$ & & $[0, t_\rho[ \, \cup \, ]t_1, \infty[$ & $]t_1, \infty[$ & $]t_\chi, \infty[$ & 
\hspace{-20mm}  
%\    
\\[2mm] \hline 
\\[-3.8mm] \hline
\end{tabular}
\end{table*}

%%%%%%%%%%%%%%%%%%%%%%

\subsection{Thermodynamic schemes}
\label{subsec-scheme-ideal}

Now that we have analyzed the hydrodynamic constraints for physical reality for the ideal case, those regarding the hydrodynamic quantities, we can proceed to analyze the thermodynamic ones. In order to do so, the thermodynamic schemes of this particular subset of solutions must be obtained. This can easily be done by substituting (\ref{g-ideal}) and (\ref{beta(alpha)-ideal}) in (\ref{n-general}) and (\ref{temperatura-general}), and using (\ref{p-ideal}). It is worth remarking that in these particular cases, the metric function $\alpha(r)$ is a function of state given by (\ref{alpha(rho,p)-1}). Furthermore, we have that $\dot{\alpha} = 0$ and, therefore, it is a particular solution of $u(s) = 0$. 

Taking all this into account, we have that the ideal Szafron model has a specific entropy $s$ which is an arbitrary function of the function of state $\alpha(\rho,p)$ given in  (\ref{alpha(rho,p)-1}), $s(\rho,p) = s(\alpha)$. Moreover, they have a matter density given by
\begin{equation} \label{n(rho,p)-ideal-1}
n(\rho,p)\! =\! \frac{1}{N(\alpha)[c\alpha^2 t^{1\!+\! \sigma}\! + \!  (1\!+\!c) \alpha t \!+\! t^{1-\sigma}]}  , 
\end{equation}
and the temperature $\Theta(\rho,p)$ is given by (\ref{temperatura-general-a}), with
\begin{subequations} \label{temperatura-ideal}
\begin{eqnarray}
	\ t_1(t) \equiv \frac{1\!-\!\sigma}{2} t^{-(1\!+\!\sigma)}, \quad \  \tau_1(r) \equiv \frac34 N'/s' ,   \qquad  \qquad \quad  \  \qquad  \\[-0mm]
	\  t_2(t) \equiv  \frac{1\!+\!\sigma}{2} t^{-(1\!-\!\sigma)}  , \quad \  \tau_2(r) \equiv \frac34 c (N \alpha^2)'/s' , \quad  \qquad  \qquad   \\[0mm]  
	\  t_3(t) \equiv \frac{1\!-\!\sigma^2}{2} t^{-1}, \quad \quad \,  \tau_3(r) \equiv \frac34 (1\!+\!c)(N \alpha)'/s' ,  \,  \ \quad  \qquad
\end{eqnarray}
\end{subequations}
where $N(\alpha)$ is an arbitrary function of $\alpha(\rho,p)$, and $t=t(p) \equiv \sqrt{(1\!-\!\sigma^2)/(3p)}$.

The set of thermodynamic schemes $\{n, s, \Theta\}$ associated with each ideal Szafron model presented above defines a family of fluids that gives the different interpretation of the solutions, and solves the so-called inverse problem for this case \cite{CFS-LTE}. One of these schemes corresponds to the generic ideal gas, namely, the one that fulfills the ideal gas equation of state (\ref{gas-ideal}). 

In \cite{CFS-LTE} we have given an algorithm to obtain this ideal gas thermodynamic scheme from the indicatrix function $\chi(\pi)$ (see also \cite{CFS-RSS}). This algorithm involves determining two quadratures which cannot be computed for the indicatrix function (\ref{chi(pi)-1}). Nevertheless, we can alternatively look for the functions $s(\alpha)$ and $N(\alpha)$ that lead to this ideal scheme. Indeed, in this case the temperature $\Theta$ and the specific energy density $e = \rho/n$ depend on $\pi$: $\Theta = \Theta(\pi)$, $e=e(\pi)$. Then, if we impose these conditions on the expressions of $n(\rho,p)$ given in (\ref{n(rho,p)-ideal-1}) and of $\Theta(\rho,p)$ given in (\ref{temperatura-general}, \ref{temperatura-ideal}), we obtain the ideal gas thermodynamic schemes if
\be
N(\alpha) = n_1 \alpha^{\frac{1-\sigma}{\sigma}}  , \qquad s(\alpha) = s_0 + s_1 \alpha^2  ,
\ee
where $s_1 = 9 \tilde{k}/(16 \sigma)$. The positivity conditions P given in (\ref{P}) and the compressibility conditions H$_2$ given in (\ref{H2-Theta}) must be required for each thermodynamic scheme to define a physically realistic fluid. For the ideal gas scheme, conditions P hold if we take $n_1 <0$. Moreover, H$_2$ can be stated in terms of the indicatrix function as \cite{CFS-LTE}
\be \label{H2-chi}
\chi(\pi) > \xi(\pi) \equiv \frac{\pi}{2 \pi +1} \,  .
\ee
Note that $\chi(0) = \chi'(0) = 0$, $\xi(0) = 0$, $\xi'(0) = 1$. Consequently, condition (\ref{H2-chi}) does not hold in a neighborhood of zero. Nevertheless, for the three cases considered in Fig. \ref{fig-1} the indicatrix function $\chi(\pi)$ fulfills this constraint if $\pi > \pi_m > 0$. 

%%%%%%%%%%%%%%%%%%%%%%%%%

%%%%%%%%%%%%%%%%%%%%%%%%%

\section{The ideal model $f(t)=\sqrt{t}$} 
\label{sec-t^1/2}

Now we consider $q=1/2$, that is $f(t)=\sqrt{t}$. Then, the solution of the field equations takes form (\ref{sol. general eq. Z}) with
\begin{equation} \label{g-ideal-2}
f(t)=\sqrt{t}  , \qquad g(t) = \sqrt{t} \ln t   \, .
\end{equation}
From these expressions we now obtain that the only functions $T_i =T_i(t)$ that do not vanish as
\begin{equation}
\label{eq:T-Szafron-2}
T_2 = -T_3 = \frac18  t^{-6} . \qquad
\end{equation}
Thus, the ideal sonic condition (\ref{ISC}) holds when $R_2 = R_3$.  The expressions (\ref{eq:R}) for these functions imply that the functions $\alpha(r)$ and $\beta(r)$ fulfill the following relation,
\begin{equation} \label{beta(alpha)-ideal-2}
\beta(\alpha) =  \alpha + \tilde{c}\, ,  
\end{equation}
where $\tilde{c} \not= 0$ is a constant. Then, taking into account that (\ref{beta(alpha)-ideal-2}) is a differential equation that relates $a(r)$ and $b(r)$, we obtain that a solution of the perfect fluid Einstein equations which is compatible with the ideal sonic condition {\rm S}$^\textrm{G}$ (\ref{chi-gas-ideal}) is given by the metric {\rm (\ref{metric-Y})} with the following election of the metric function $Y(t,r)$,
\begin{equation} \label{Y-ideal}
Y=Z^{2/3} , \quad      Z(t,r) =  b(r)\sqrt{t} [\ln t + \alpha(r)]  , 
\end{equation}
where $b(r)$ is given by
\begin{equation} \label{b-ideal}
b(r) = e^{\alpha(r)/\tilde{c}} \,  ,  \quad \tilde{c}\not=0 \,  .
\end{equation}
%
%
%Moreover, we have $g_{rr}= Y'^2$, with
%
%\be
%Y' = \frac{2\,t^{\frac{1}{3}} e^{\frac{2 \alpha}{3 \tilde{c}}}}{3 \tilde{c}} \frac{\alpha'(\alpha + \tilde{c} + \ln t)}{(\alpha  + \ln t)^{\frac13}} \, .
%\ee
%

From (\ref{expansion-Y}), the expansion in this case is 
\begin{equation} \label{expansion-ideal-2} \displaystyle
\theta = \frac{1}{t}\left[ 1 + \frac{1}{\alpha + \ln t} + \frac{1}{\alpha + \tilde{c} + \ln t}\right]  \, .
\end{equation} 
Again, the change $t \leftrightarrow -t$, with $t<0$, leads to contracting models, whose properties are similar to those we study below for the expanding models.

%%%%%%%%%%%%%%%%%%%%%%%%%

\subsection{Hydrodynamic quantities: Energy density, pressure, and speed of sound} 
\label{subsec-hydro-ideal-2}

Now, the pressure also takes the expression (\ref{p-ideal}), and the energy density is
\begin{equation} \label{rho-ideal-2}
\rho =  \frac{1}{3t^2} \frac{(2 + \alpha + \ln t)(2 + \alpha + \tilde{c} + \ln t)}{(\alpha + \ln t)(\alpha + \tilde{c} + \ln t)}   .
\end{equation}
From this expression we obtain
\begin{eqnarray} \label{alpha(rho,p)-2}
\hspace{-4.0mm} 
\alpha = \tilde{\alpha} (\rho,p) \equiv  \frac{4\pi + \varepsilon \tilde{F}(\pi)}{2(1\!-\!\pi)}  + \frac12 \ln (3p)\!-\! \frac{\tilde{c}}{2} \, , \\[2mm]
\label{F(pi)b}
	\tilde{F}(\pi) \equiv  \sqrt{\tilde{c}^2(1-\pi )^2 + 16\pi}   \, .
\end{eqnarray}
Then, we can determine the indicatrix function $\chi=\chi(\pi)$,
\begin{equation} \label{chi(pi)-2}
c_s^2 = \tilde{\chi}(\pi) \equiv \frac{4\pi^2[ \tilde{c}^2 (1+\pi) + 2 \varepsilon \tilde{F}(\pi)]}{(1+\pi) [ \tilde{c}^2 (1+\pi)^2 + 16\pi]} \, . 
\end{equation}

It is worth remarking that this ideal model can be obtained from the ideal Szafron model studied in the previous section by a limit procedure taking $\sigma \rightarrow 0$, $c \rightarrow 1$ and $\hat{c} =(1-c)/\sigma \rightarrow \tilde{c}$. Consequently, to study this model we can start from the expressions obtained in this subsection, or we could sometimes use the analysis already made for the ideal Szafron models.

Now, when $t \rightarrow 0$ or when $t \rightarrow \infty$, the solution becomes a shift FLRW model $p=\rho$, as we can deduce by taking $\sigma =0$ in (\ref{limits}).

%%%%%%%%%%%%%%%%%%%%%%

\subsection{Fluid properties: Compressibility conditions {\rm H}$_1^{\rm G}$}
\label{subsec-compress-ideal-2}

We must analyze the behavior of the equation of state $\chi(\pi)$ in the interval $0 < \pi < 1$ where the energy conditions E$^\textrm{G}$ hold. Now $\chi(\pi)$ defines a family of fluids depending on the parameter $\tilde{c}^2$ and the sign $\varepsilon$.  

The function $\chi(\pi)$ can have two different behaviors depending on the sign $\varepsilon$. If we choose the positive sign, $\varepsilon=+1$, $\chi(\pi)$ is an increasing function, and the first of the compressibility conditions H$_1^{\rm G}$ holds in the whole interval $0 < \pi < 1$, and $\chi(1)=1$. This case appears as the limit of the ideal Szafron model plotted in Fig. \ref{fig-1}a. 

If $\varepsilon=-1$, $\chi(\pi)$ identically vanishes when $|\tilde{c}| =2$. When $|\tilde{c}| >2$ it is an increasing function, and the causal condition holds in the whole interval $0 < \pi < 1$. This case appears as the limit of the ideal Szafron model plotted in Fig. \ref{fig-1}c. Otherwise, when $|\tilde{c}| <2$, $\chi(\pi)$ is a negative function and it does not fulfill the causal condition for any interval. 

The second compressibility condition in H$_1^{\rm G}$ holds in the interval where $\chi(\pi)$ is well defined and fulfills the causal condition in the cases considered above.

%%%%%%%%%%%%%%%%%%%%%%%

%%%%%%%%%%%%%%%%%%%%%%

\subsection{Curvature singularities and spacetime domains} 
\label{subsec-singularities-2}

For this model we can also have up to three different singularities. First, a big bang singularity at $t = 0$. Secondly, the metric line element of the sphere vanishes and the energy density is infinite at $\tilde{t}_1 = \tilde{t}_1(r) = e^{-\alpha(r)}$. Finally, a singularity could appear at $\tilde{t}_2 = \tilde{t}_2(r) = e^{-\alpha(r)-\tilde{c}}$, and then the metric distance on the coordinate lines of coordinate $r$ vanishes and we have infinite energy density again.

Now, two different situations can be distinguished:
\begin{itemize}
\item[(i)] If $|\tilde{c}| = 2$, we have the singularity at $t = 0$ and also the one at $\hat{t}(r)$, where $\hat{t} = t_1$ if $\tilde{c} = 2$, and $\hat{t} = t_2$ if $\tilde{c} = -2$. Now, two disconnected spacetime domains exist: ${\cal R}_0 = \{0 < t < \hat{t}\}$ and $\hat{\cal R} = \{\hat{t}<  t < \infty\}$.
\item[(ii)] If $|\tilde{c}| \not=  2$, then we have all three singularities, and three disconnected spacetime domains exist: ${\cal R}_0 = \{0 < t < t_m\}$, ${\cal R}_m = \{t_m < t < t_M\}$  and ${\cal R}_M = \{t_m<  t < \infty\}$, where $t_m = {\rm min}\{t_1,t_2\}$ and $t_M = {\rm max}\{t_1,t_2\}$.
\end{itemize}
Note that these two cases are similar to cases (ii) and (iii) considered in Sec. \ref{subsec-singularities} for the ideal Szafron models (see Figs. \ref{fig-2}b and \ref{fig-2}c).

%%%%%%%%%%%%%%%%%%%%%%

\subsection{Analysis of the solutions and energy conditions} 
\label{subsec-energy-c-2}

The spacetime domains where the energy conditions E$^{\rm G}$ and the compressibility conditions H$_1^{\rm G}$ hold depend on the value of $\tilde{c}$. These domains are defined by the times $t_1(r)$ and $t_2(r)$ that determine the singularities and the times $t_\rho= t_\rho(r) = e^{-\alpha(r) - 2}$, $\tilde{t}_\rho =\tilde{t}_\rho(r) = e^{-\alpha(r) - \tilde{c} - 2}$ and $t_\pi = t_\pi(r) = e^{-(2 + 2\alpha(r) + \tilde{c})/2}$. 

We have a behavior that is similar to some cases of the ideal Szafron models summarized in Table \ref{table-1}. If $|\tilde{c}|=2$ (respectively, $|\tilde{c}|<2$ or $|\tilde{c}|>2$) the behavior is that of the fifth and sixth rows (respectively, third and fourth rows) in Table \ref{table-1}. Exchanging $\tilde{c}  \leftrightarrow  -\tilde{c}$ produces the exchange of $t_1  \leftrightarrow  t_2$ and $t_\rho  \leftrightarrow  \tilde{t}_\rho$.

%%%%%%%%%%%%%%%%%%%%%%

\subsection{Thermodynamic schemes}
\label{subsec-scheme-ideal-2}

The specific entropy $s$ is again an arbitrary function of the function of state $\alpha(\rho,p)$ (\ref{alpha(rho,p)-2}), $s(\rho,p) = s(\alpha)$. Moreover, the matter density is given by
\begin{equation} \label{n(rho,p)-ideal}
n(\rho,p) = \frac{4\sqrt{3p}}{N(\alpha)[2\alpha - \ln(3p)][2\alpha + 2\tilde{c} - \ln(3p)]} \, , 
\end{equation}
where $N(\alpha)$ is an arbitrary function of $\alpha(\rho,p)$. And the temperature $\Theta(\rho,p)$ is given by (\ref{temperatura-general-a}), with 
\begin{subequations} \label{temperatura-ideal-2}
\begin{eqnarray}
	\ t_1(t) \equiv \frac{1}{t}(1 + \ln t)  , \qquad  \, \tau_1(r) \equiv \frac34 N'/s' ,   \qquad  \qquad \quad  \ \ \qquad  \\[-0mm]
	\  t_2(t) \equiv  \frac{1}{2t}   , \qquad \qquad \quad \  \ \tau_2(r) \equiv \frac34 [N \alpha (\alpha + \tilde{c})]'/s' ,  \quad  \qquad \
	  \\[0mm]  
	\  t_3(t) \equiv \frac{1}{2t}(1 +\frac12  \ln t)  , \quad   \tau_3(r) \equiv \frac34 [N (2 \alpha+ \tilde{c})]'/s' ,  \,  \ \quad  \qquad
\end{eqnarray}
\end{subequations}
where $t=t(p) \equiv 1/\sqrt{3p}$.

In this case, the ideal thermodynamic scheme that fulfills the equation of state (\ref{gas-ideal}) can be obtained by taking $N(\alpha) = n_0 e^{- \alpha}$ and $s(\alpha) = s_0 + s_1 \alpha$, with $n_0>0$ and $s_1 = -9\tilde{k}/8$.

%%%%%%%%%%%%%%%%%%%%%%%%%

\section{Open topics and work in progress}
\label{sec-progress}

Our study on the thermodynamics of the spatially flat LT metrics (\ref{metric-Y}) analyzes several significant issues. Nevertheless, there are some open problems that are not solved yet and that require endeavor beyond the scope of this paper. Now, we present some preliminary results of them.

%%%%%%%%%%%%%%%%%%%%%%%%%

\subsection{On the models with homogeneous temperature}
\label{subsec-T(t)}

According to the thermodynamic theory of irreversible processes (in both the standard irreversible thermodynamics \cite{Eckart} and the extended irreversible thermodynamics \cite{Israel-76, Israel-St}), the transport coefficients of thermal conductivity, shear-viscosity, and bulk-viscosity appear in the constitutive equations linking dissipative fluxes (anisotropic pressures, bulk viscous pressure, and energy flux) with the kinematic coefficients of fluid flow (shear, expansion and acceleration) \cite{Rezzolla}.  

The perfect fluid approximation can be considered when the transport coefficients of a fluid vanish (or are negligible). A non-perfect fluid is a fluid with at least a non-zero transport coefficient. For this fluid, the energetic evolution is, generically, described by an energy tensor with energy flux and anisotropic pressures. However, when a non-perfect fluid admits particular evolutions in which the dissipative fluxes vanish, these evolutions are well described by a perfect energy tensor, and are usually called {\em equilibrium states} \cite{Rezzolla}. Moreover, all the thermodynamic relations of the perfect fluid hydrodynamics remain valid. Furthermore, the shear, the expansion and the acceleration of the fluid undergo strong restrictions as a consequence of the constitutive equations. Specifically, if the thermal conductivity coefficient does not vanish, then the fluid acceleration is constrained by the relation:
\be \label{Fourier}
a = -\! \perp \! \dif \ln \Theta \, , 
\ee
where $\perp$ denotes the orthogonal projection to the fluid velocity.

After these considerations, we can look for perfect fluid solutions to the Einstein equations that describe both (i) a thermodynamic perfect fluid in local thermal equilibrium, and (ii)  an inviscid (with negligible shear and bulk viscosity coefficients) non-perfect fluid in equilibrium. Then, the thermal conductivity coefficient does not vanish and, when the fluid flux is geodesic as the solutions we are considering here, equation (\ref{Fourier}) implies a homogeneous temperature $\Theta= \Theta(t)$.

Thus, a forthcoming study we can address is to find the solutions with homogeneous temperature. To do so, we must analyze the compatibility of the expression of the temperature (\ref{temperatura-general}), with the constraint $\Theta= \Theta(t) \not=0$. This analysis requires us to consider different cases. 

For example, we can look for the thermodynamic schemes $\{s(\alpha), N(\alpha)\}$ and the metric function $\beta(\alpha)$ which are compatible with a homogeneous temperature for any solution $\{f(t), g(t)\}$ of the field equations. In this case the three functions $\tau_i(r)$ given in (\ref{temperatura-general}) are constant and then a straightforward calculation leads to:
\begin{subequations} \label{T(t)-beta}
\begin{eqnarray}
\beta(\alpha) =  \frac{\beta_0 \alpha + \beta_1}{\beta_2 \alpha - \beta_0}   ,  \qquad  \label{T(t)-beta-a}\\[0mm]
s(\alpha) = \frac{n_0 \alpha \beta(\alpha) - s_1}{s_0 - n_1 \alpha \beta(\alpha)}, \quad  N(\alpha) = n_0\! +\! n_1 s(\alpha) ,  \qquad   \label{T(t)-beta-b} 
\end{eqnarray}
\end{subequations}
with $\beta_0 = n_1 s_1 - n_0 s_0$. Note that $\beta_1$ and $\beta_2$ cannot be canceled simultaneously (this leads to $s'(\alpha)=0$). Thus, the above expression of $\beta(\alpha)$ is compatible with neither (\ref{beta(alpha)-ideal}) nor (\ref{beta(alpha)-ideal-2}). Moreover, the three functions $t_i(t)$ given in (\ref{temperatura-general}) are independent for the ideal models studied in Secs. \ref{sec-t^q} and \ref{sec-t^1/2}. Consequently, these models cannot represent an inviscid fluid with a non-vanishing thermal conductivity coefficient. The compatibility of (\ref{T(t)-beta}) with the general ideal sonic condition (\ref{ISC}) leads to a system of five fourth-order differential equations, which will be studied elsewhere.

When at least one of the functions $\tau_i(r)$ is non-constant we can consider different cases that lead to solutions admitting thermodynamic schemes with homogeneous temperature. They will be analyzed elsewhere.

%%%%%%%%%%%%%%%%%%%%%%%%%

\subsection{On the solutions of the ideal sonic condition}
\label{subsec-ISC}

As commented in Sec. \ref{subsec-idealvaris} the study of the general solution of the ideal sonic equation (\ref{ISC}) is a task that falls outside the scope of this work. A way to extend the family of solutions is to consider a constraint for the function $\beta(\alpha)$, compute the functions $R_i(\alpha)$ given in (\ref{eq:R}), and analyze the subsequent equation (\ref{ISC}) for $\{f(t), g(t)\}$.

As an example, let us consider $\beta = -\alpha$. Then, $R_1 = R_4 = R_5 = R_8 =0$, $R_2 = -R_3 = 2 \alpha$, $R_7 = -R_6 = 2 \alpha^3$. Consequently, Eq. (\ref{ISC}) is equivalent to
\begin{subequations} \label{R1=0}
\begin{eqnarray}
T_2 - T_3 \equiv E_1(\ddddot{f}, \dddot{f}, \ddot{f},\dot{f}, f, g) = 0 \, ,   \label{R1=0-a}\\[1mm]
T_6 - T_7 \equiv E_2(\ddddot{f}, \dddot{f}, \ddot{f},\dot{f}, f, g) = 0 \, .   \label{R1=0-b} 
\end{eqnarray}
\end{subequations}
Then, we can eliminate the fourth derivative from (\ref{R1=0}) and obtain (considering a non-constant pressure): 
\be  \label{R1=0-2}
E_3 \equiv \frac{\dddot{f}}{\ddot{f}} \frac{\dot{g}}{g} 
- \frac{\ddot{f}}{\dot{f}} \left[\frac{\dot{f}}{f}+\frac{\dot{g}}{g}\right] + \frac{\dot{g}^2}{g^2}= 0 \, .
\ee
It is easy to prove that this equation implies (\ref{R1=0}). Thus, the functions $\{f(t), g(t)\}$ must fulfill the third-order differential system (\ref{gt(f)}, \ref{R1=0-2}), and then a solution for each initial condition $\{f(t_0), \dot{f}(t_0), \ddot{f}(t_0), g(t_0)\}$ exists. It is worth remarking that the ideal Szafron models with $c=-1$ are the solution to these equations for specific initial conditions. The study of the solutions corresponding to other initial conditions requires a numerical approach that is beyond the scope of this work.

%%%%%%%%%%%%%%%%%%%%%%%%%

\subsection{On the evolution of the energy density profiles}
\label{subsec-EEDP}
For each choice of the functions $\alpha(r)$, expressions (\ref{rho-ideal-1}) and (\ref{rho-ideal-2}) give different energy density profiles that could
model inhomogeneities. The evolution of these inhomogeneities will be studied in forthcoming works.
Actually, a similar analysis was carried out for some subclasses of the general solution (\ref{def. Z})-(\ref{sol. general eq. Z}) in the context of the so-called "Swiss cheese” models \cite{Bona-Stela-c}. Moreover, one of these subclasses seem to admit solutions compatible with the ideal sonic condition (\ref{ISC}). Therefore, the study of the thermodynamics of such solutions and their possible interpretation as a generic ideal gas are open problems that we leave for future work.

%%%%%%%%%%%%%%%%%%%%%%%%%%%%%%

\begin{acknowledgements}
This work has been supported by the Spanish Ministerio de Ciencia e Innovaci\'on and the Fondo Europeo de Desarrollo Regional, Projects PID2019-109753GB-C21 and PID2019-109753GB-C22 and the Generalitat Valenciana Project AICO/2020/125. 
\end{acknowledgements}

%%%%%%%%%%%%%%%%%%%%%%%%%%

%\nocite{*}
\bibliography{PRD}
% Produces the bibliography via BibTeX.

\end{document}